\documentclass[preprint]{aastex}

\begin{document}


\title{Sensitivity of astrophysical reaction rates to nuclear uncertainties}


\author{T. Rauscher}
\affil{Department of Physics, University of Basel, CH-4056 Basel, Switzerland}


%\altaffiltext{1}{Visiting Astronomer, Cerro Tololo Inter-American Observatory.
%CTIO is operated by AURA, Inc.\ under contract to the National Science
%\altaffiltext{2}{Society of Fellows, Harvard University.}
%\altaffiltext{3}{present address: Center for Astrophysics,
%    60 Garden Street, Cambridge, MA 02138}
%\altaffiltext{4}{Visiting Programmer, Space Telescope Science Institute}
%\altaffiltext{5}{Patron, Alonso's Bar and Grill}


\begin{abstract}
Sensitivities of nuclear reaction rates to a variation of nuclear properties are studied. Target nuclei range from proton- to neutron-dripline for $10\leq Z\leq 83$. Reactions considered are nucleon- and $\alpha$-induced reactions mediated by the strong interaction. The contribution of reactions proceeding on the target ground state to the total stellar rate is also given. General dependences on various input quantities are discussed. Additionally, sensitivities of laboratory cross sections of nucleon-, $\alpha$-, and $\gamma$-induced reactions are shown, allowing to estimate the impact of cross section measurements. Finally, recommended procedures to explore and improve reaction rate uncertainties using the present sensitivity data are outlined.
\end{abstract}


\keywords{astrophysical reaction rates --- nucleosynthesis --- compound nucleus reaction}



\section{Introduction}
\label{sec:intro}

Nuclear reaction rates are a central quantity in all investigations making use of reaction networks to
study nucleosynthesis and energy generation by nuclear reactions in astrophysical environments.
The majority of reaction rates appearing in astrophysical plasmas has to be theoretically predicted due to a number of reasons.
Reactions on highly unstable nuclei, mostly appearing in explosive nucleosynthesis, cannot be accessed in the laboratory, yet. Close to and at stability, tiny reaction cross sections due to Coulomb barriers often prevent a measurement at astrophysically relevant energies. Finally, even if experimental reaction cross sections are available, they are only known for nuclei in their ground states. Due to thermal excitation in the plasma, however, a considerable fraction of nuclei may be in excited states and reactions proceeding on those have to be included in the calculation of the reaction rate.

A frequently encountered question in this context is the one asking for a quantification of the uncertainties in the predictions. The knowledge of allowed variations in a given reaction rate may be useful to also determine the connected astrophysical uncertainties or to even rule out a specific scenario. The question, however, is difficult to address because theoretical uncertainties are fundamentally different from the uncertainty quantification generally used as "error bar" in measurements. It is possible, however, to identify different factors contributing to the "theoretical uncertainty" of a rate and to discuss them separately.

The first factor is the choice of the theoretical framework, the model. Obviously, it is \textit{a priori} impossible to exactly determine the correctness of a model without being able to falsify it through comparison with data. Even previous knowledge to justify the assumptions can only be used to motivate the choice of a model but not strictly prove its validity. This impossibility poses the fundamental problem of quantifying uncertainties of the theory approach. It is loosely related to systematic errors in experiments due to instruments and external effects, which are also hard to estimate. Systematic errors in measurements can usually be removed after identifying the cause. For theory, this would mean to give up a model for another one, driven by the failure to describe data. This is not just of philosophical importance. As discussed below, an important implication of this is the fact that model limitations and applicability can only be estimated and the estimate cannot be strictly quantified in a general manner.

Assuming that a "correct" model is used, the next step in the uncertainty analysis would be to check the reliability of the input required to perform calculations with the model. The input is comprised of fundamental constants, such as the speed of light or the value of the Planck constant, and other quantities, e.g., nuclear masses or other nuclear properties. Part of this input is known experimentally and carries well defined statistical errors. These errors can be propagated in the model, either through systematic variation within the given limits or by a Monte Carlo analysis, and their impact on the final prediction can be quantified. Unfortunately, there are two further classes of input, for which this is not possible. One of these classes is experimental input bearing a large systematic uncertainty. Specifically for our purposes here, this may include incomplete nuclear level schemes, especially from nuclear spectroscopy of unstable nuclides. Finally, some of the required input may again be provided by theoretical models. This will be the case for the majority of the input required to predict reaction rates for nuclei far off stability. This last input class suffers from the same fundamental problem we encounter when trying to quantify uncertainties for our model of choice. Therefore it is impossible to assign an "error bar" to this input type and know its impact on the final prediction. This cannot be simply remedied by comparing results of different theoretical models. This would not be a systematic variation covering the range of possible values. A set of models may coincidentally yield very similar values for a given quantity (even if only in the energy range relevant for the calculation of a specific rate) but may be equally incorrect.

The impact of a variation in the input on the final result can be studied independently of the actual uncertainties. Of course, this still involves the assumption of having chosen an appropriate model to correctly describe the physical situation. Although this will not lead to a full quantification of the uncertainties, it provides important information in showing which input quantities have the largest contribution to the overall uncertainty in the prediction. This allows to focus on improving the uncertainties in the most important quantities and to rule out certain other quantities to have an effect on the predicted value. Such an approach also has the advantage that it does not distinguish between uncertainties stemming from experiment and from theory. Having made a choice (or guess) of uncertainties in the input, they can be simply plugged into the sensitivities of the model to arrive at a final error estimate.

This work aims at providing tables of the sensitivities of nuclear reaction rate predictions for nucleon- and $\alpha$-induced reactions mediated by the strong interaction. The considered target nuclei range from proton- to neutron-dripline for $10\leq Z\leq 83$. In addition, sensitivities of laboratory reaction cross sections to a variation of the same properties are given. This will aid the conception of experiments as well as theoretical investigations to reduce the uncertainties in the most important properties.

\section{Definitions}

\subsection{Sensitivity}
\label{sec:sensidef}

In order to quantify the impact of a variation of a model quantity $q$ (directly taken from input or derived from it) on the final result $\Omega$, the relative sensitivity $^\Omega s_q$ is defined as \citep{raureview}
\begin{equation}
\label{eq:sensi}
^\Omega s_q=\frac{v_\Omega-1}{v_q-1} \quad.
\end{equation}
It is a measure of a change by a factor of $v_\Omega=\Omega_\mathrm{new}/\Omega_\mathrm{old}$ in $\Omega$ as the result of a change in the quantity $q$ by the factor $v_q=q_\mathrm{new}/q_\mathrm{old}$, with $^\Omega s_q=0$ when no change occurs and $^\Omega s_q=1$ when the final result changes by the same factor as used in the variation of $q$, i.e., $^\Omega s_q=1$ implies $v_\Omega=v_q$.

The sign of the sensitivity $^\Omega s_q$ provides further information on the change and is also given in the tables presented in \S~\ref{sec:results}. Since both $v_\Omega>0$ and $v_q>0$ for the quantities studied in this work, a positive sign implies that $\Omega$ changes in the same manner as $q$, i.e., $\Omega$ becomes larger when the value of the quantity $q$ is increased. The opposite is true for $^\Omega s_q<0$, i.e., $\Omega$ decreases with an increase of $q$.

The above definitions are consistent with the ones used in standard sensitivity analysis when realizing that Equation (\ref{eq:sensi}) is equivalent to
\begin{equation}
^\Omega s_q=\frac{\frac{d\Omega}{\Omega_\mathrm{old}}}{\frac{dq}{q_\mathrm{old}}} \quad,
\end{equation}
with $d\Omega=\Omega_\mathrm{new}-\Omega_\mathrm{old}$ and $dq=q_\mathrm{new}-q_\mathrm{old}$.

In the following, the final result $\Omega$ is either a rate (r) or a reaction cross section (cs). The varied quantities are (averaged) widths and the nuclear level density.

\subsection{Reaction rates}
\label{sec:ratedef}

The stellar reaction rate $r^*$ is calculated by folding a Maxwell-Boltzmann distribution $\Phi_\mathrm{MB}(E,T)$ of the relative interaction energies between target $A$ and projectile $a$ with the effective cross section $\sigma^\mathrm{eff}$ which describes the probability that a reaction occurs under the given plasma conditions \citep{fow74,hwfz,raureview}
\begin{equation}
\label{eq:stellarrate}
r^*=\frac{n_a n_A}{G_0(T)} \int_0^\infty \sigma^\mathrm{eff}(E) \,
\Phi_\mathrm{MB}(E,T) \, dE = n_a n_A R^* \quad.
\end{equation}
For simplicity, the reactivity $R^*$ (or rate per particle pair) is also called rate in some publications.
Here, the energy $E$ is measured relative to the ground state (g.s.) of $A$. The plasma conditions are given by the number densities $n_a$, $n_A$ of projectiles and targets, and by the temperature $T$. The normalized partition function $G_0$ of the target accounts for the effect of thermally populated excited states,
\begin{equation}
G_0=\frac{\sum_\mu (2J_\mu+1)\exp\left(-E_\mu/(kT)\right)}{2J_0+1} \quad,
\end{equation}
with $J_\mu$ being the spins of ground and excited states, and $E_\mu$ their excitation energies relative to the ground state. (The g.s.\ is assigned $\mu=0$ and $E_0=0$.)

The effective cross section \citep{fow74,wardfowler,hwfz,raureview} includes a weighted sum over all possible transition cross sections $\sigma^{\mu \rightarrow \nu}$ connecting all initial states $\mu$ in nucleus $A$ at $E_\mu \leq E$ with all energetically accessible final states $\nu$ in nucleus $B$, i.e., with all states with $E_\nu \leq E+Q$, where $Q$ is the reaction $Q$ value,
\begin{equation}
\sigma^\mathrm{eff}=\sum_\mu \sum_\nu \frac{2J_\mu+1}{2J_0+1} \frac{E-E_\mu}{E}
\sigma^{\mu \rightarrow \nu}(E-E_\mu) \quad.
\end{equation}
Following \citet{fow74}, the above equation implicitly assumes that cross sections $\sigma^{\mu \rightarrow \nu}$ at zero or negative energies $E-E_\mu$ are Zero.

The effective cross section differs from the definition of the laboratory cross section $\sigma^\mathrm{lab}=\sum_\nu \sigma^{0 \rightarrow \nu}$ in that it also includes weighted contributions from excited states in the target. This accounts for the thermal population of excited states in an astrophysical plasma. Their contribution is more important in nuclei with a larger number of low-lying excited states and/or at elevated temperature. The g.s.\ contribution $X$ to the stellar rate,
\begin{equation}
\label{eq:xfactor}
X(T)=\frac{r^\mathrm{lab}}{r^* G_0(T)}=\frac{\int_0^\infty \sigma^\mathrm{lab}(E) \,
\Phi_\mathrm{MB}(E,T) \, dE}{\int_0^\infty \sigma^\mathrm{eff}(E) \,
\Phi_\mathrm{MB}(E,T) \, dE} \quad,
\end{equation}
provides a measure for the importance of excited states \citep{xfactor}. It can only assume values in the range $0\leq X\leq 1$, where the laboratory (ground state) cross section is sufficient to calculate the stellar rate when $X=1$. A smaller $X$ shows increasing contribution of excited state transitions to the stellar rate and diminishing importance of the g.s.\ transitions. For example, $X=0.3$ implies that 70\% of the stellar rate is determined by transitions from excited states in the target nucleus and only 30\% of the rate stems from reactions on the g.s. The value of $X$ is also a measure by how much a theoretical uncertainty can be reduced by a measurement of the laboratory cross section (for a further discussion, see \S~\ref{sec:examples}).

Stellar rates, but \textit{not} laboratory rates, obey an important reciprocity relation connecting forward and reverse reaction \citep{fow74,wardfowler}. The stellar reactivity $R^*_{AB}$ of the reaction $a+A \rightarrow B+b$ is connected to the reactivity $R^*_{BA}$ of the inverse reaction $b+B \rightarrow A+a$ by
\begin{mathletters}
\begin{eqnarray}
\frac{R^*_{BA}}{R^*_{AB}} &\propto & e^{-Q_{AB}/(kT)} \quad, \label{eq:recipart}\\
\frac{R^*_{BA}}{R^*_{AB}} &\propto & T^{3/2} e^{-Q_{AB}/(kT)} \quad. \label{eq:reciphoto}
\end{eqnarray}
\end{mathletters}
Equation (\ref{eq:recipart}) applies when projectile $a$ and ejectile $b$ are particles and Equation (\ref{eq:reciphoto}) applies when projectile $a$ is a particle and $b$ is a photon. The above relations were derived from the well-known reciprocity theorem of nuclear reactions \citep[see, e.g.,][]{BW52} by considering that $\sigma^\mathrm{eff}$ includes transitions to all energetically accessible states in the target and residual nucleus.

\section{Reaction models}
\label{sec:reacmodels}

Here we only want to outline the relevant features of the models to be able to discuss their sensitivities. For a detailed discussion of reaction models used in astrophysics, see, e.g., \citet{De03,descrau,raureview}.

The majority of nuclear reaction rates were calculated in the compound nucleus reaction model, applying the statistical model by \citet{hf52}, see, e.g., \citet{hwfz,whfz,sar82,ctt,nonsmoker,rath,most,cyburt}. The statistical model assumes a sufficiently high nuclear level density (NLD) in the compound nucleus $C$ formed by a reaction $a+A\rightarrow C \rightarrow B+b$, with target $A$, projectile $a$, residual nucleus $B$, and ejectile $b$. In the case of capture reactions, the ejectiles are photons. The central assumption in this model is that the compound nucleus is formed at an excitation energy where such a large number of resonances of all required spins and parities is encountered that they cannot be individually resolved. Instead of decay widths of an isolated resonance, the widths of "average" resonances are used. It can be shown that the statistical model can be derived from the single-resonance Breit-Wigner formula by averaging over a large number of resonances \citep{descrau}.

\begin{figure}
%\plottwo{f2.eps}{f2_color.eps}
\includegraphics[width=\columnwidth]{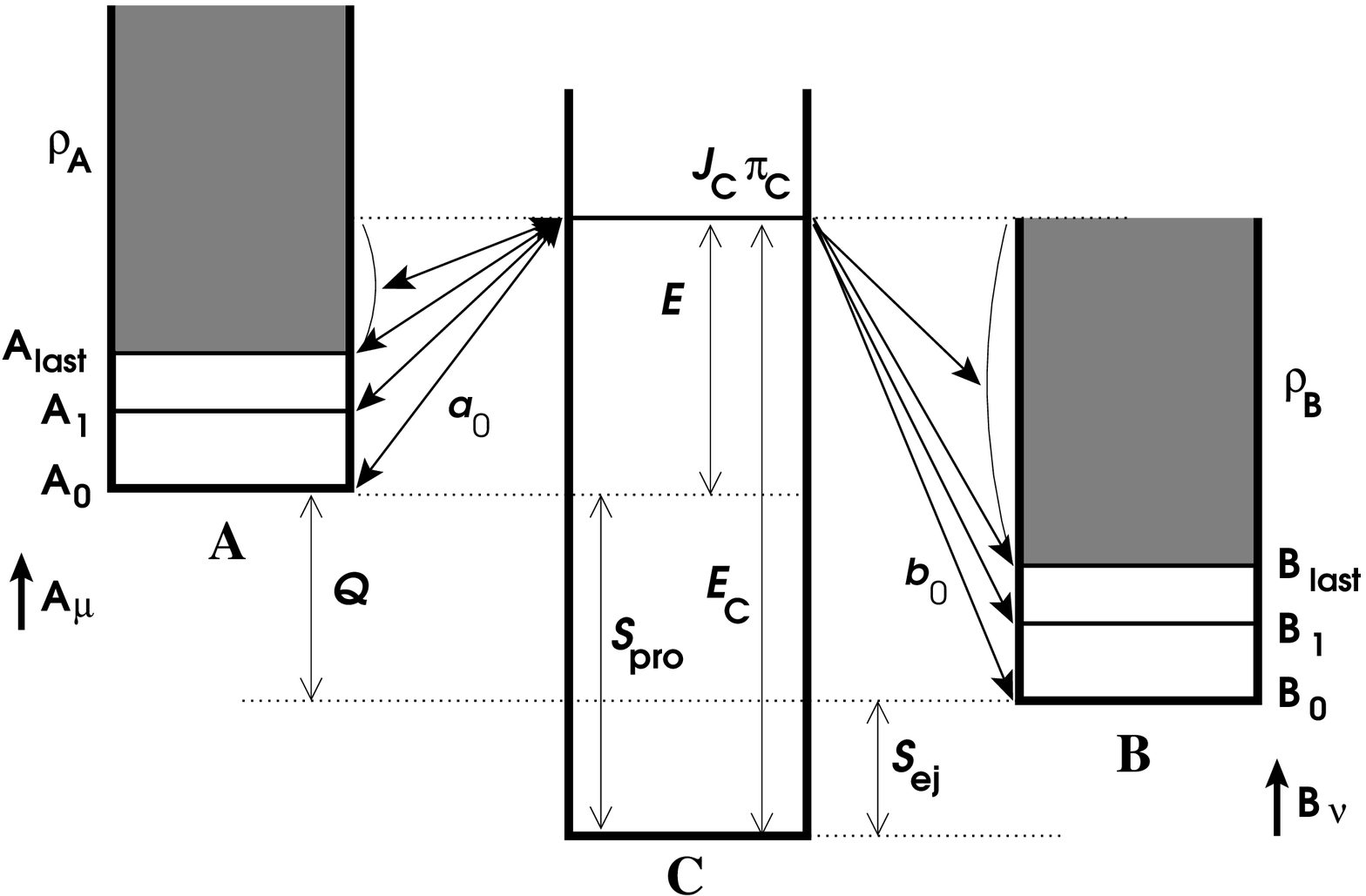}
\caption{Energy scheme of a compound reaction $a+A\rightarrow C \rightarrow B+b$, proceeding via a compound state at energy $E_\mathrm{C}$ with spin $J_\mathrm{C}$ and parity $\pi_\mathrm{C}$. Transitions to and from states $A_\mu$ in the target nucleus and $B_\nu$ in the final nucleus are shown. In a capture reaction, the "ejectile" is a photon and nuclei $B$ and $C$ coincide. In a laboratory reaction, only the transition from the g.s.\ ($a_0$) contributes in $A$ whereas all transitions to and from the g.s.\ and all excited states of $A$ and $B$ with $E_{\mu,\nu} \leq E_\mathrm{C}-S_\mathrm{pro,ej}$, respectively, contribute in the effective cross section $\sigma^\mathrm{eff}$.\label{fig:transitions}}
\end{figure}

The energy scheme for a compound reaction is shown in Figure \ref{fig:transitions}. It applies to both the case of isolated resonances and the statistical model. The latter would additionally involve a sum over many compound resonances $J_\mathrm{C}$, $\pi_\mathrm{C}$ at each $E_\mathrm{C}$ whereas there are only few, well separated resonances with discrete $E_\mathrm{C}$ in the other case.

The sensitivities of rates and cross sections presented in this work were computed in the statistical model with version 0.8.3 of the SMARAGD code \citep{smaragd,raureview}. Nevertheless, similar sensitivities would be expected when the used widths were not averaged widths of a large number of compound resonances but of a single, narrow resonance dominating the cross section. The rates depend on similar expressions in both cases, a fraction
\begin{equation}
\label{eq:f}
\mathcal{F}(E_C,J_C,\pi_C)=(2J_C+1)\frac{W_{E_C,J_C,\pi_C}^A W_{E_C,J_C,\pi_C}^B}{W_{E_C,J_C,\pi_C}^\mathrm{all}} \quad.
\end{equation}
The quantities $W^A$, $W^B$, and $W^\mathrm{all}$ are the total width of a compound level in $C$ with spin $J_C$ at excitation energy $E_C$ and including all transitions to levels $\mu$ in nucleus $A$ with $E_\mu \leq E_C-S_\mathrm{pro}$ in the entrance channel, the total width of this compound state including all possible transitions to levels $\nu$ in $B$ with $E_\nu \leq E_C-S_\mathrm{ej}$ in the exit channel, and the combined width $W_{E_C,J_C,\pi_C}^\mathrm{all}=\sum_{\lambda=A, B, \dots} W_{E_C,J_C,\pi_C}^\lambda$ from the sum of all emission processes from the compound state to all channels $\lambda$, respectively \citep{hwfz,raureview}. The separation energies of the projectile and the ejectile in the compound nucleus $C$ are denoted by $S_\mathrm{pro}$ and $S_\mathrm{ej}$, respectively. Assuming narrow resonances, the rate of a single Breit-Wigner resonance at excitation energy $E_C$ (this would be an energy of $E=E_C-S_\mathrm{pro}$ in the entrance channel) is given by \citep{fcz67,raureview}
\begin{equation}
\label{eq:BWrate}
r_\mathrm{BW}^* \propto \frac{1}{G_0(kT)^{3/2}} e^{\left( E_C-S_\mathrm{pro} \right) /(kT)} \mathcal{F}(E_C,J_C,\pi_C) \quad.
\end{equation}
In the Hauser-Feshbach model it is assumed that there is a large number of resonances at any compound formation energy $E_C$ and for any $J_C$ and parity $\pi_C$. Therefore, the rate is
\begin{equation}
\label{eq:HFrate}
r_\mathrm{HF}^* \propto \frac{1}{G_0(kT)^{3/2}} \int_0^\infty \sum_{J_C,\pi_C} \mathcal{F}(E+S_\mathrm{pro},J_C,\pi_C) \Phi_\mathrm{MB}(E,T) dE\quad.
\end{equation}
Although the sum over $J_C$ is unrestricted, only few values will actually contribute, due to quantum mechanical selection rules and the fact that contributions of large partial waves are strongly suppressed by the angular momentum barrier.

Expressions for the laboratory rates are simply obtained by dropping the factor $1/G_0$ in Equations (\ref{eq:BWrate}) and (\ref{eq:HFrate}) and by using only one transition in the determination of the $W^A$ appearing in the numerator of Equation (\ref{eq:f}), the one to the g.s.\ of $A$ \citep{hwfz,raureview}.

The importance of direct reactions -- and specifically of direct neutron capture -- in astrophysical environments has also been discussed in literature \citep[see, e.g.,][and references therein]{MaMe83,microdirect,gordc,raureview}. These are not included in the present sensitivity study. In Figure \ref{fig:transitions} they would be shown as positive-energy transitions directly between thermally populated states in the nuclei $A$ and $B$, without the formation of an intermediate compound state in $C$. The energy difference between initial and final states would be emitted in a single photon. Their sensitivities, however, are trivial to estimate because only single transitions are involved instead of summed widths. Thus, the rate of a direct reaction will linearly depend on the spectroscopic factors used, with a sensitivity $s=1$. They will also be more strongly sensitive to a change in the properties of the single final state (excitation energy, spin, parity) than compound reactions \citep{microdirect}.

Direct reactions at high energies -- (d,p) and (d,n) reactions at several tens to hundreds of MeV, for instance -- have also been suggested to be used in nuclear spectroscopy studies of unstable nuclei at radioactive ion beam facilities. Cross sections of reactions at such high energies do not appear in the calculation of astrophysical reaction rates but the extracted information on low-lying excited states and spectroscopic factors is useful in the calculation of the widths $W$ implicitly appearing in Equations (\ref{eq:BWrate}) and (\ref{eq:HFrate}) (see also \S~\ref{sec:dependence}).

\subsection{Model applicability}
\label{sec:applicability}

As pointed out above, the statistical model relies on the assumption that a large number of narrow compound resonances is present at the formation energy of the compound nucleus. Whether this is valid needs to be ascertained, in principle, for each reaction and incident projectile energy separately. As this would involve a comparison of the actual number and widths of resonances to the averaged widths predicted by the model, this is not strictly possible within the model itself. This is the first type of fundamental uncertainty described in \S~\ref{sec:intro}. It has to be mentioned, however, that there is a difference in the applicability of the Hauser-Feshbach model to reaction cross sections and to astrophysical reaction rates, as these involve different energy distributions of the projectiles and therefore average in a different manner. In an experimental determination of a cross section the energy distribution of the projectiles is given by the energy distribution within the original particle beam and the energy straggling within the target. Considerable efforts are usually made to confine the resulting energies to a narrow range, in order to limit the energy uncertainty in the measured cross section. Therefore, resonances with a width comparable to the obtained energy resolution will be noticeable in the excitation function. In the calculation of a reaction rate, on the other hand, the cross section is averaged over a Maxwell-Boltzmann energy distribution (see Equation \ref{eq:stellarrate}) which is much wider than the energy resolution of an experiment. This explains the fact that the reaction rate is rather "forgiving" to deviations around the "true" cross section value, provided the deviations cancel within the integration. In consequence, the Hauser-Feshbach model is more widely applicable for reaction rates than for cross sections. It will even be applicable in the presence of narrow resonances within the astrophysically important energy range as long as their average contribution within this range is correctly accounted for.

Applicability maps for neutron-, proton-, and $\alpha$-induced reactions were shown in \citet{rtk97}, specifying the lowest stellar temperature at which the statistical model is estimated to be still suitable to predict the stellar rate. (It should be noted that these were estimates based on a predicted NLD.) The temperature limits were also explicitly included in the tables of \citet{rath}. The rates for the majority of reactions with $Z \gtrsim 10$ can be described within the statistical model. The exceptions are rates for which the dominant contributions to the reaction rate integral stem from compound excitation energies with low NLD. For neutron-induced reactions this can be the case even close to and at stability for closed shell nuclei. The Gamow windows are shifted to higher compound excitation in reactions for which the energy dependence is determined by a charged-particle width. Therefore these are not so sensitive to closed shells at stability. Reactions on heavy nuclei, however, may involve $\alpha$-induced reactions with negative $Q$ value. This may move the Gamow window back into a region of low NLD. Moreover, proton- and neutron-induced reactions close to the proton- and neutron dripline, respectively, also show low or negative $Q$ values and also for these cases the statistical model cross sections may be suited only to calculate rates at high temperature.

The astrophysically relevant energy windows as function of temperature have recently been revised by \citet{energywindows}. This impacts the derived applicability limits only in a limited manner, as explained in \S~5.3.2 of \citet{raureview}.

\section{Results and Discussion}
\label{sec:results}

\subsection{Sensitivity tables for width variations}
\label{sec:sensitables}

\subsubsection{Rates and ground state contributions}
\label{sec:ratetables}

Table \ref{tab:sensi} contains the sensitivities (as defined in Equation \ref{eq:sensi}) for astrophysical reaction rates when systematically and separately varying the neutron-, proton-, $\alpha$-, and $\gamma$-widths by a factor $v_q=2$. These widths are always included in $W^\mathrm{all}$ of Equation (\ref{eq:HFrate}) and, depending on the reaction studied, can also appear as $W^\mathrm{ent}$ or $W^\mathrm{ex}$. The g.s.\ contribution $X$ to the stellar rate is also given. The table is organized as follows: The first column contains the abbreviated element name of the target nucleus $A$ and the next two columns give its charge $Z_A$ and mass number $M_A$. The fourth and fifth columns identify the reaction by specifying projectile $a$ and ejectile $b$, respectively. Roman letters are used in the identification of projectile and ejectile: n (neutron), p (proton), a ($\alpha$), g ($\gamma$). This is followed by 24 blocks for the 24 temperatures 0.1, 0.15, 0.2, 0.3, 0.4, 0.5, 0.6, 0.7, 0.8, 0.9, 1.0, 1.5, 2.0, 2.5, 3.0, 3.5, 4.0, 4.5, 5.0, 6.0, 7.0, 8.0, 9.0, and 10.0 GK. Each block contains five columns showing the sensitivity $^\mathrm{r} s_q$ to a variation of the $\gamma$-, neutron-, proton-, and $\alpha$-width ($^\mathrm{r} s_\mathrm{g}$, $^\mathrm{r} s_\mathrm{n}$, $^\mathrm{r} s_\mathrm{p}$, $^\mathrm{r} s_\mathrm{a}$, in that order) and the g.s.\ contribution $X$ (Equation \ref{eq:xfactor}) at the given temperature. The table is sorted by $Z$, $A$, and reaction type.

\begin{figure}
%\plottwo{f2.eps}{f2_color.eps}
\includegraphics[angle=-90,width=\columnwidth]{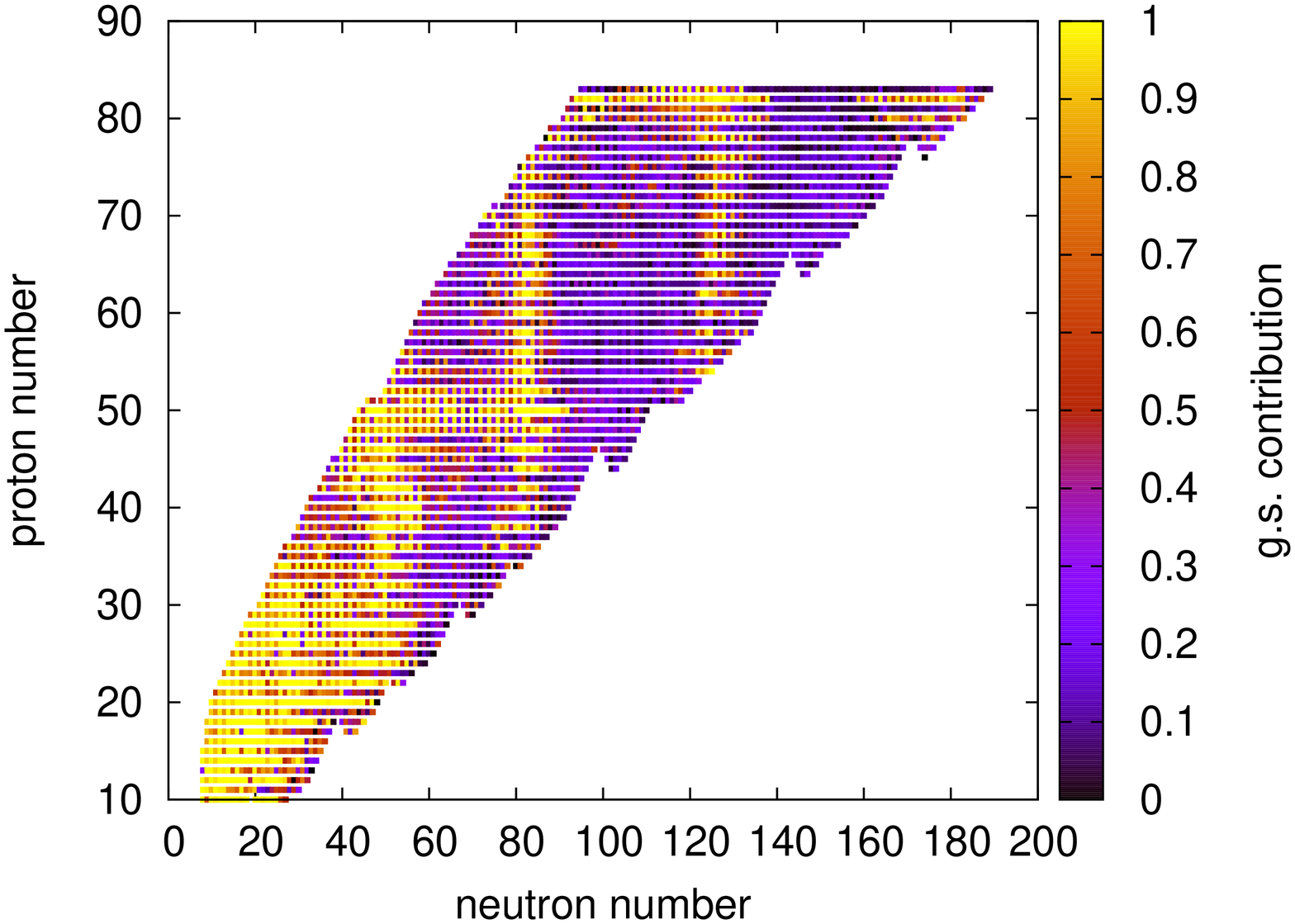}
\caption{Ground state contribution $X$ to stellar (n,$\gamma$) rates at 1.5 GK. See the electronic edition of the Journal for a color version of this figure.\label{fig:Xng1p5}}
\end{figure}

\begin{figure}
%\plottwo{f2.eps}{f2_color.eps}
\includegraphics[angle=-90,width=\columnwidth]{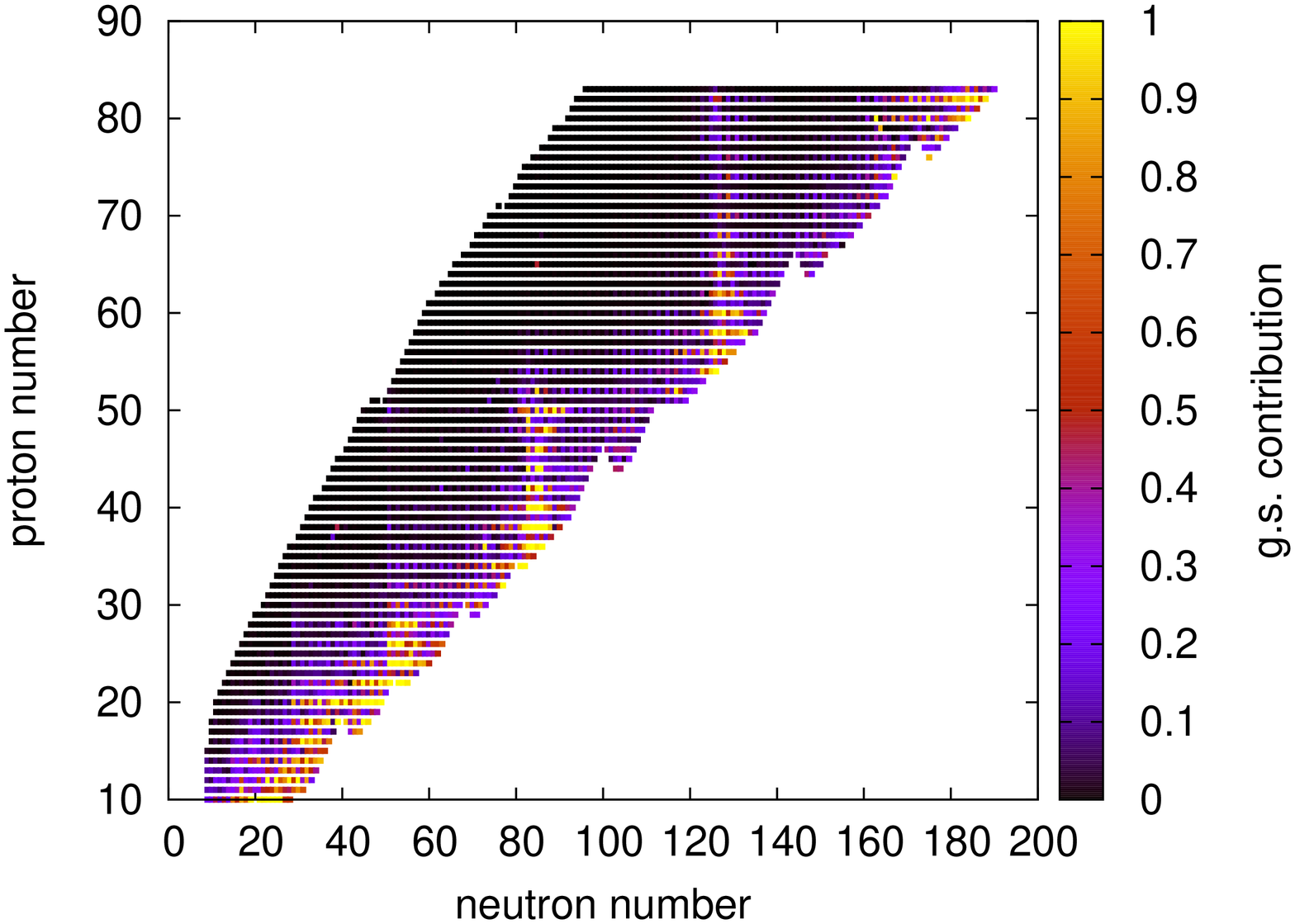}
\caption{Ground state contribution $X$ to stellar ($\gamma$,n) rates at 1.5 GK. See the electronic edition of the Journal for a color version of this figure.\label{fig:Xgn1p5}}
\end{figure}

\begin{figure}
%\plottwo{f2.eps}{f2_color.eps}
\includegraphics[angle=-90,width=\columnwidth]{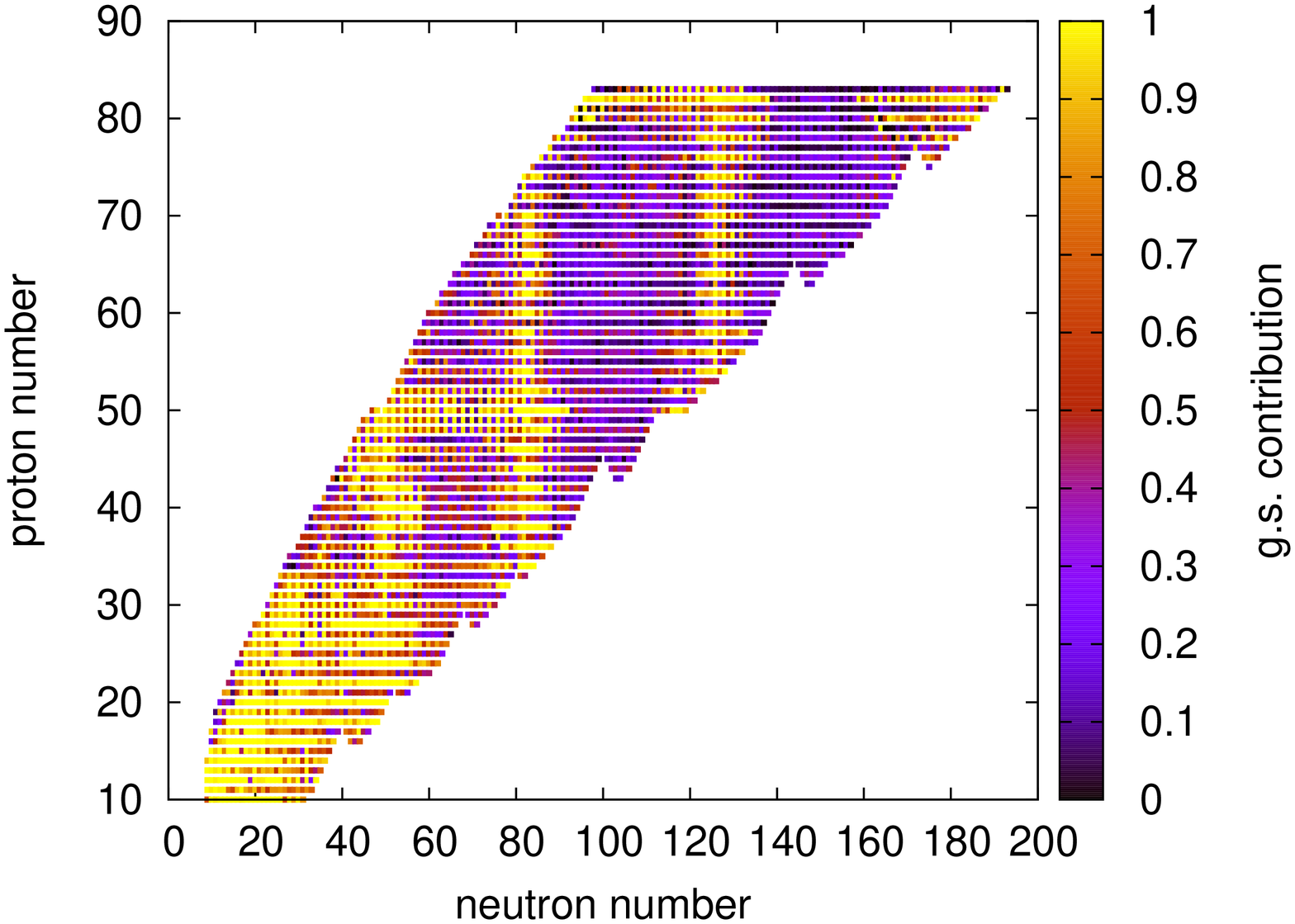}
\caption{Ground state contribution $X$ to stellar (p,$\gamma$) rates at 1.5 GK. See the electronic edition of the Journal for a color version of this figure.\label{fig:Xpg1p5}}
\end{figure}

\begin{figure}
%\plottwo{f2.eps}{f2_color.eps}
\includegraphics[angle=-90,width=\columnwidth]{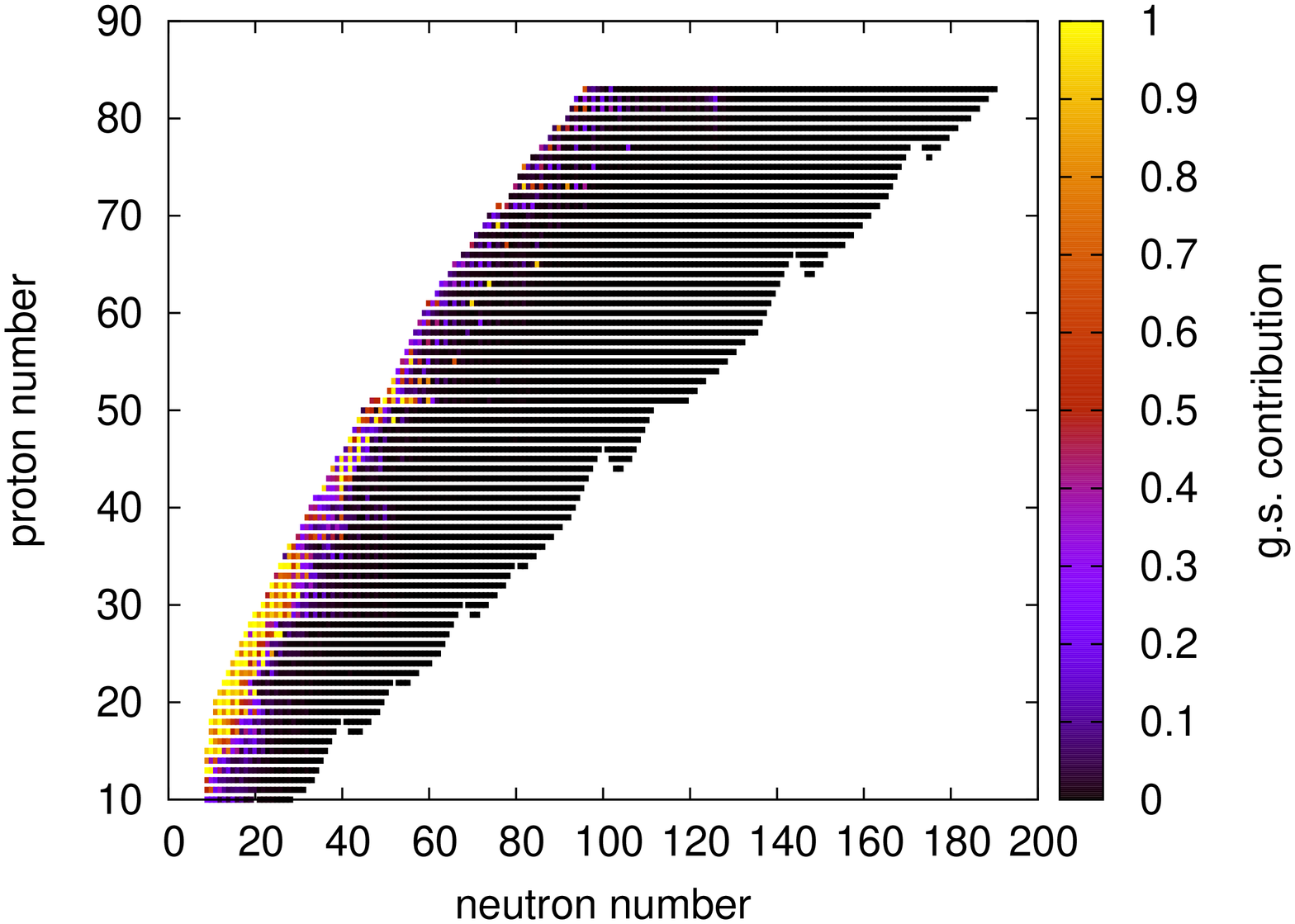}
\caption{Ground state contribution $X$ to stellar ($\gamma$,p) rates at 1.5 GK. See the electronic edition of the Journal for a color version of this figure.\label{fig:Xgp1p5}}
\end{figure}

\begin{figure}
%\plottwo{f2.eps}{f2_color.eps}
\includegraphics[angle=-90,width=\columnwidth]{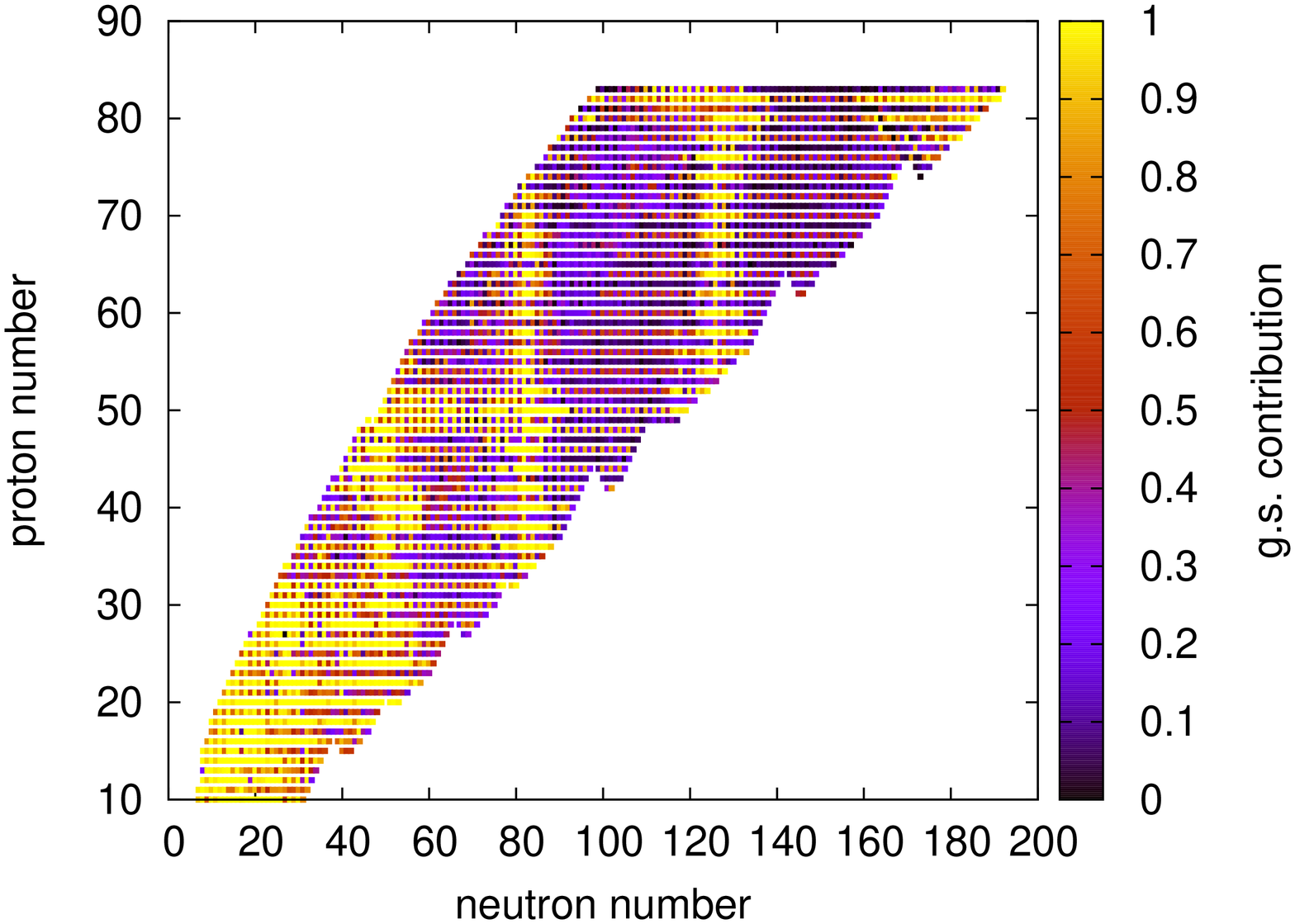}
\caption{Ground state contribution $X$ to stellar ($\alpha$,$\gamma$) rates at 1.5 GK. See the electronic edition of the Journal for a color version of this figure.\label{fig:Xag1p5}}
\end{figure}

\begin{figure}
%\plottwo{f2.eps}{f2_color.eps}
\includegraphics[angle=-90,width=\columnwidth]{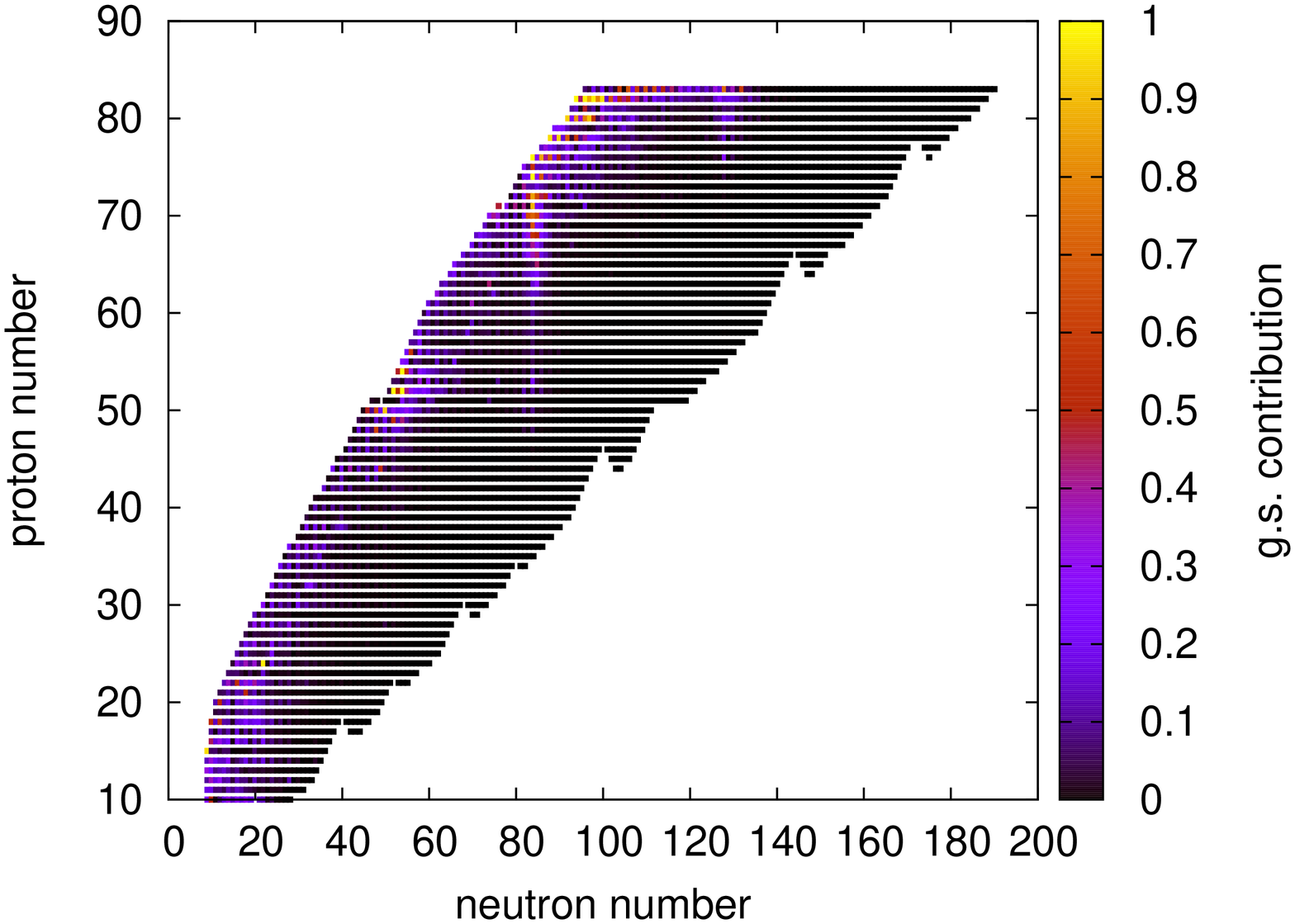}
\caption{Ground state contribution $X$ to stellar ($\gamma$,$\alpha$) rates at 1.5 GK. See the electronic edition of the Journal for a color version of this figure.\label{fig:Xga1p5}}
\end{figure}

Except for captures, only reactions with positive reaction $Q$ value are shown. Due to the reciprocity of stellar rates (Equations \ref{eq:recipart}, \ref{eq:reciphoto}), the sensitivities are the same for forward and reverse reaction. The g.s.\ contributions, on the other hand, can be quite different. For the majority of rates, $X$ is smaller for the reaction direction with negative $Q$ value. Only comparatively few exceptions are known to this rule. They are due to a Coulomb suppression effect of excited state contributions which is discussed in \citet{kisscoul,raucoul}. Because it may be of interest for experimental nuclear physics, the g.s.\ contributions $X$ for reactions with negative $Q$ value are given in Table \ref{tab:xnegq}. A similar table (Table \ref{tab:xphoto}) is given for photodisintegration rates. This is mainly to illustrate that especially photodisintegration rates have much smaller g.s.\ contributions than the respective capture rates. In consequence, the determination of a photodisintegration cross section on a target nucleus in the g.s.\ does not provide much information on the stellar rate. Moreover, it exhibits a different sensitivity to the input as can be seen from the tables provided in \S~\ref{sec:cstables}. To provide a quick overview, Figures \ref{fig:Xng1p5}--\ref{fig:Xga1p5} show $X$ at 1.5 GK for captures and photodisintegrations involving neutrons, protons, and $\alpha$-particles. The g.s.\ contributions will be smaller at higher temperatures. Values of $X$ for (n,$\gamma$) close to stability at s-process temperatures have been discussed in \citet{xfactor}.

The general behavior of the sensitivities can be understood by recalling a few simple facts. First, it has to be noted that in astrophysical applications forward and backward rates are in equilibrium at temperatures above 5-6 GK. Above this temperature, individual rates are not important and the attained equilibrium abundances are governed by the reciprocity relations in Equations (\ref{eq:recipart}) and (\ref{eq:reciphoto}). The temperature dependence of the sensitivity of a reaction rate to a given width is comparatively weak below 5 GK. This is because the relevant energy range covered in the integration for the rate (Equation \ref{eq:stellarrate}) is comparatively narrow as it depends on the width of the Gamow peak at the given temperature and the effective cross section is integrated over this range. It is even narrower for neutron channels where the relevant energy range is given by the width of the Maxwell-Boltzmann distribution.

As for the dependence of the rates on certain widths, a few cases can be distinguished regarding the ratio of widths in Equation (\ref{eq:f}):
\begin{enumerate}
\item The fraction $\mathcal{F}$ foremost depends on the smallest width appearing in the numerator when the value of the denominator is dominated by the larger width in the numerator. This is frequently the case as the denominator contains the sum of all channels including the ones appearing in the numerator. In this case the denominator cancels with the larger width and the smaller width remains. Its energy dependence then determines the energy dependence of the cross section. The rate and cross section will then be completely insensitive to changes in the larger width (as long as it does not become comparable or smaller than the other width in the numerator) and any change in the smaller width will transfer fully to the cross section and rate.
\item When the denominator is not dominated by either of the two widths appearing in the numerator, it is irrelevant whether they are equal or of different magnitude. The cross section and rate will be sensitive to a change in any of the widths, either one in the numerator \textit{and} a third width dominating the denominator. A change in the denominator, however, will affect the rate in inverse proportionality, i.e., an increase in a width in the denominator will lead to a decrease in the cross section and rate.
\item The most complicated case appears when the two widths in the numerator are comparable and both are strongly contributing to the total width in the denominator. The cross section and rate will then be sensitive only to changes in these two widths but the level of sensitivity depends on the size of the variation factor. The sensitivity will be low with a small variation factor because any change in the numerator is cancelled by a similar change in the denominator. Using a large variation factor, on the other hand, may transform this case to case 1 above, when the varied width becomes much larger or smaller than the other ones.
\end{enumerate}

Case 1, of course, is well known in resonance analyses performed in experimental nuclear physics. A closer inspection of the sensitivities, however, shows that a rule-of-thumb often applied in nuclear reaction analysis does not always work for astrophysical rates. The rule states that the $\gamma$-width is smaller than the particle widths and therefore the energy dependence and sensitivity of a capture rate will be solely governed by the $\gamma$-width, according to case 1 above. Astrophysical charged particle capture on intermediate and heavy nuclei (with $Z_A > 20$), however, shows a different behavior. This is due to the fact that the interaction energies of the charged particles are close to or below the Coulomb barrier and consequentially the widths are suppressed. They become comparable or smaller than the $\gamma$-width and thus determine the sensitivity behavior. This implies that $\gamma$-widths play a minor role in certain astrophysical charged-particle capture reactions, contrary to the case of standard nuclear reactions at higher energy. They still play a role in neutron captures, though, unless the relevant energy window is close to the neutron threshold. Close to the threshold, the neutron width may become smaller than the $\gamma$-width again.

\begin{figure}
%\plottwo{f2.eps}{f2_color.eps}
\includegraphics[angle=-90,width=\columnwidth]{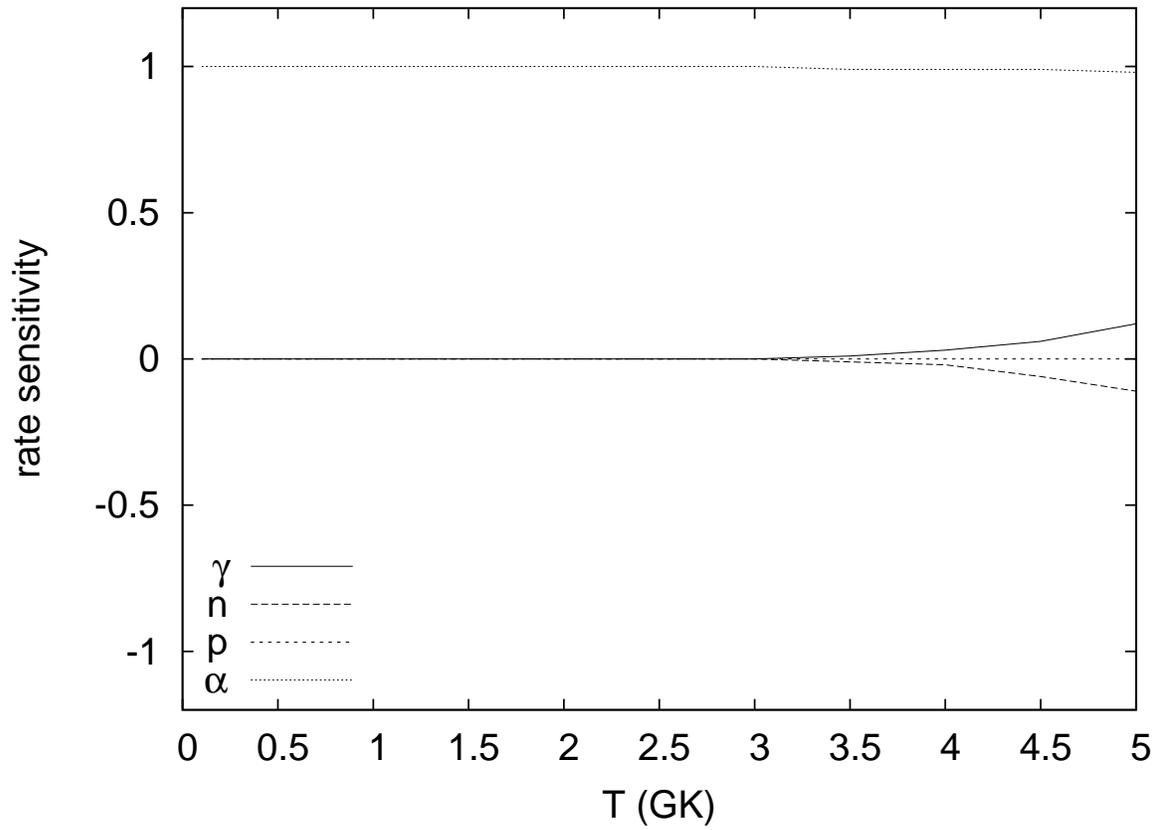}
\caption{Sensitivity of the stellar rate of $^{144}$Sm($\alpha$,$\gamma$)$^{148}$Gd to a variation of various widths as function of plasma temperature.\label{fig:rate_sm144ag}}
\end{figure}

\begin{figure}
%\plottwo{f2.eps}{f2_color.eps}
\includegraphics[angle=-90,width=\columnwidth]{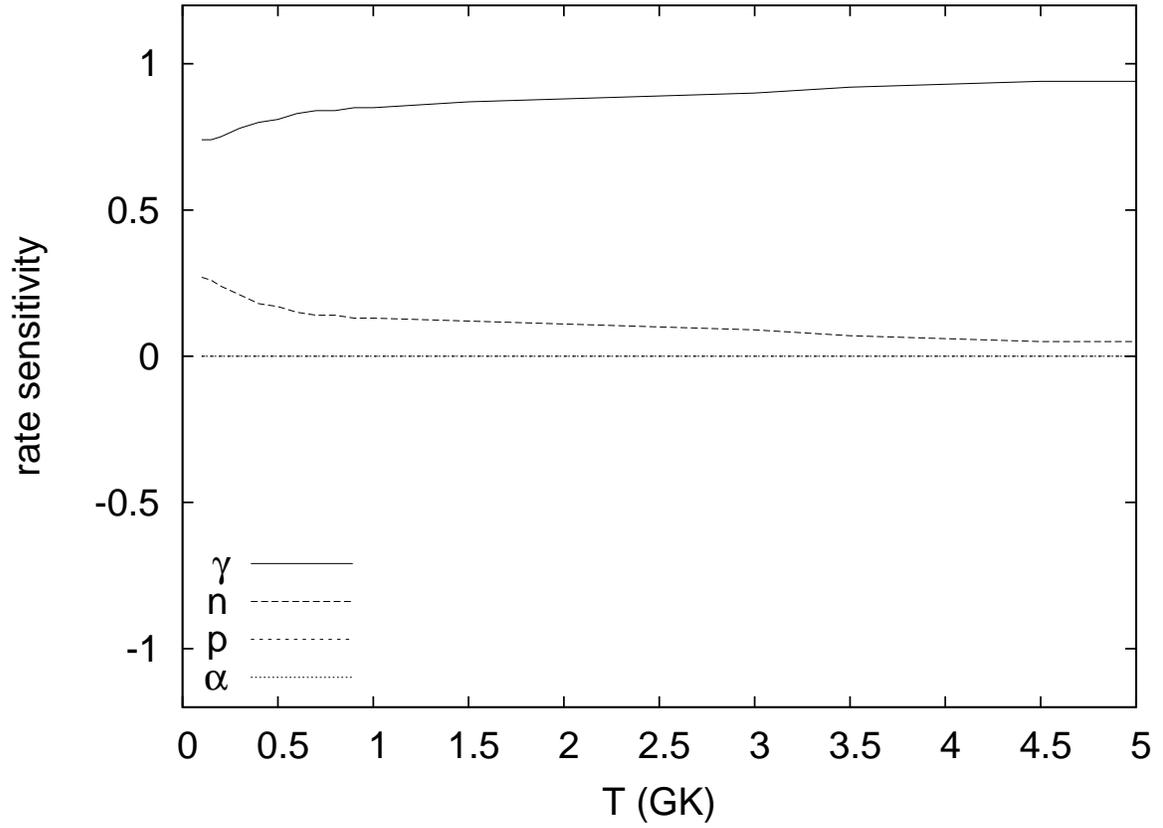}
\caption{Sensitivity of the stellar rate of $^{132}$Sn(n,$\gamma$)$^{133}$Sn to a variation of various widths as function of plasma temperature. There is zero sensitivity to both proton and $\alpha$ widths and their lines overlap.\label{fig:rate_sn132ng}}
\end{figure}

Figures \ref{fig:rate_sm144ag} and \ref{fig:rate_sn132ng} illustrate the above by showing the sensitivities on different width variations of the stellar rates for $^{144}$Sm($\alpha$,$\gamma$)$^{148}$Gd and $^{132}$Sn(n,$\gamma$)$^{133}$Sn, respectively. While the $\alpha$ capture reaction is only sensitive to the $\alpha$-width which remains smaller than the $\gamma$-width across all astrophysically relevant temperatures and energies, the neutron capture reaction is indeed governed by the $\gamma$-width. It should be noted that the neutron capture $Q$-value of the doubly magic nucleus $^{132}$Sn is small and therefore the NLD at the compound formation energy may be too small to allow application of the statistical Hauser-Feshbach model (see \S~\ref{sec:applicability}). Single resonances and direct capture may provide important contributions to the reaction rate.

\begin{figure}
%\plottwo{f2.eps}{f2_color.eps}
\includegraphics[angle=-90,width=\columnwidth]{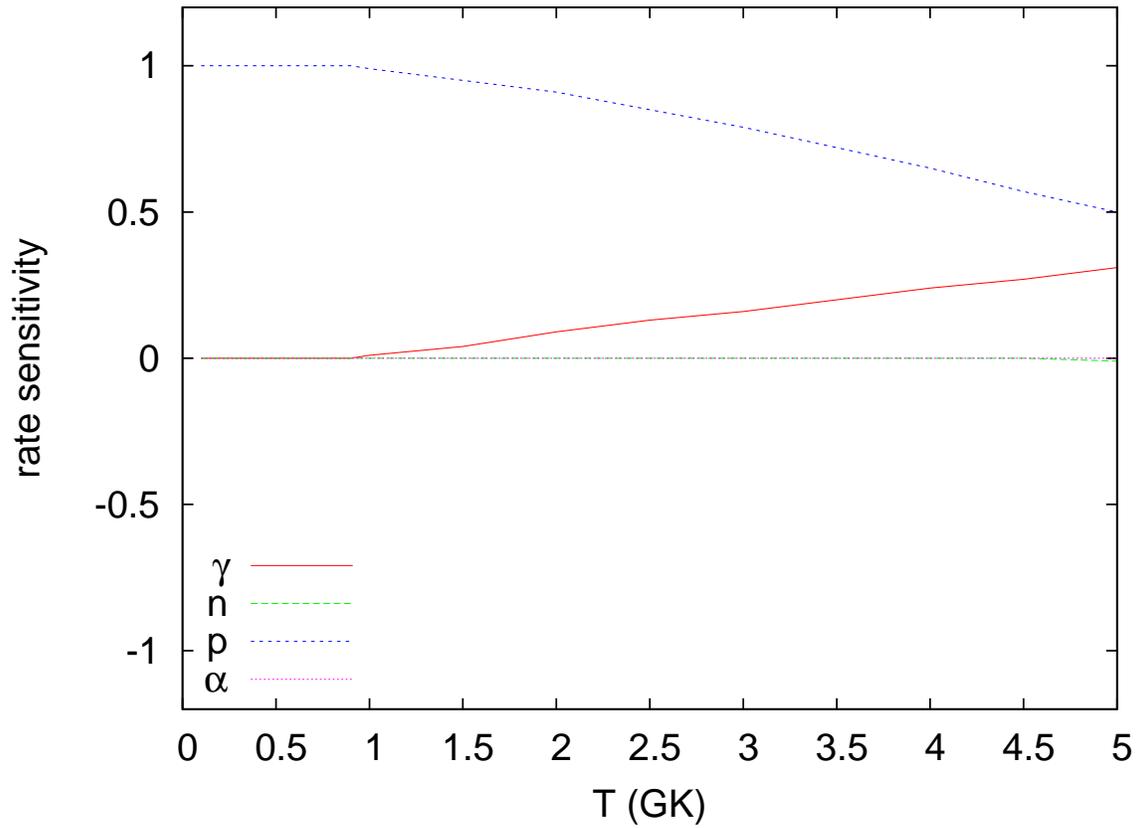}
\caption{Sensitivity of the stellar rate of $^{96}$Ru(p,$\gamma$)$^{97}$Rh to a variation of various widths as function of plasma temperature. The sensitivities to changes in both neutron and $\alpha$ widths are negligible and their lines overlap. See the electronic edition of the Journal for a color version of this figure.\label{fig:ru96rate}}
\end{figure}

As another example, the sensitivities of the stellar $^{96}$Ru(p,$\gamma$)$^{97}$Rh rate are shown in Figure \ref{fig:ru96rate}. The rate is mostly sensitive to the proton width over all of the relevant temperature range. Towards higher temperatures, however, the sensitivity to the proton width is slowly decreasing while the one to the $\gamma$ width is increasing. At 5 GK both sensitivities already are quite similar: the sensitivity to the proton width is only $^r s _p=0.5$ while the one to the $\gamma$ width is already $^r s _\gamma=0.3$. Uncertainties in other widths do not play a role.

\begin{figure}
%\plottwo{f2.eps}{f2_color.eps}
\includegraphics[angle=-90,width=\columnwidth]{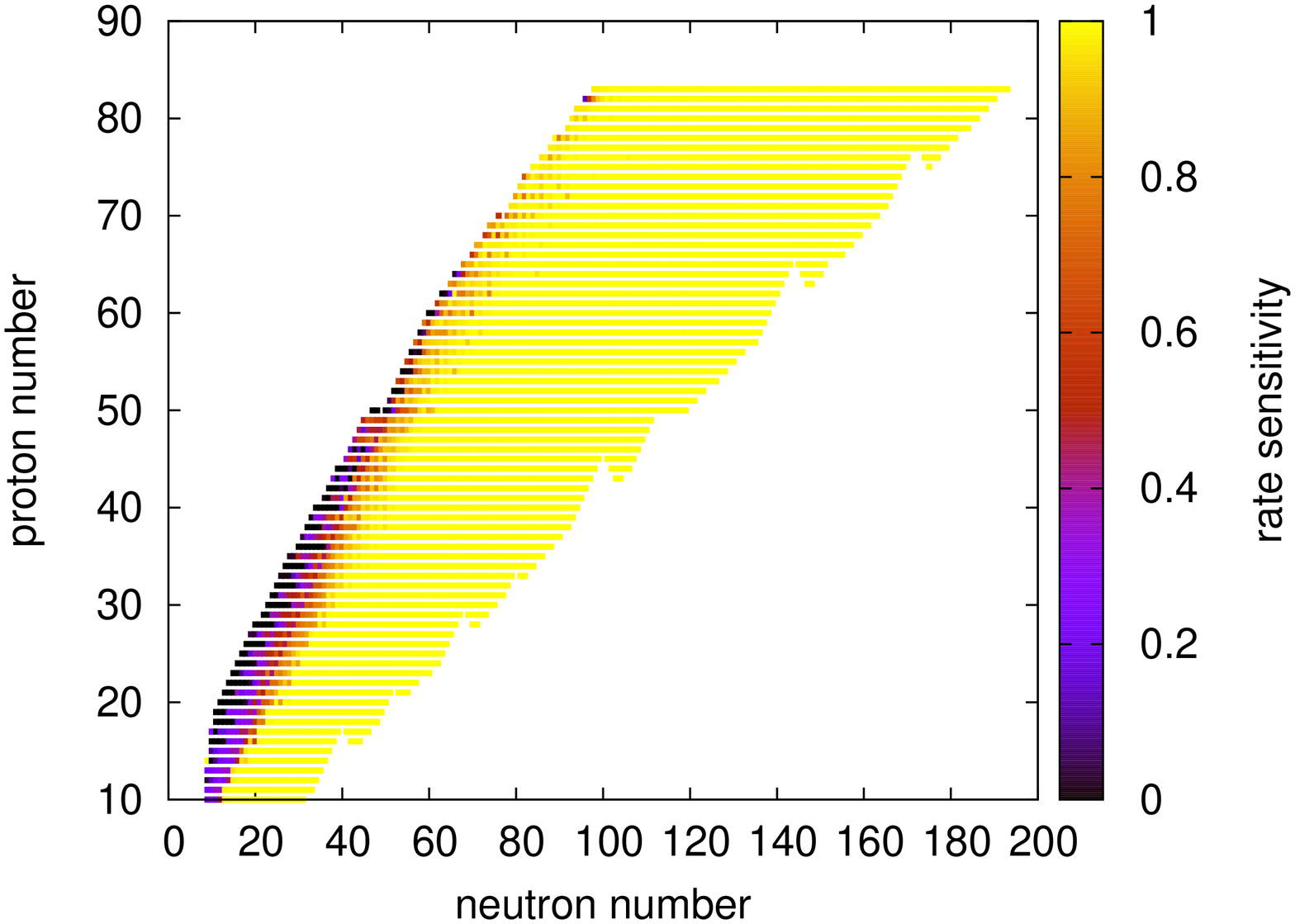}
\caption{Sensitivities of stellar proton capture rates at 1.5 GK to a variation of the proton width. See the electronic edition of the Journal for a color version of this figure.\label{fig:3Dpgprot}}
\end{figure}

\begin{figure}
%\plottwo{f2.eps}{f2_color.eps}
\includegraphics[angle=-90,width=\columnwidth]{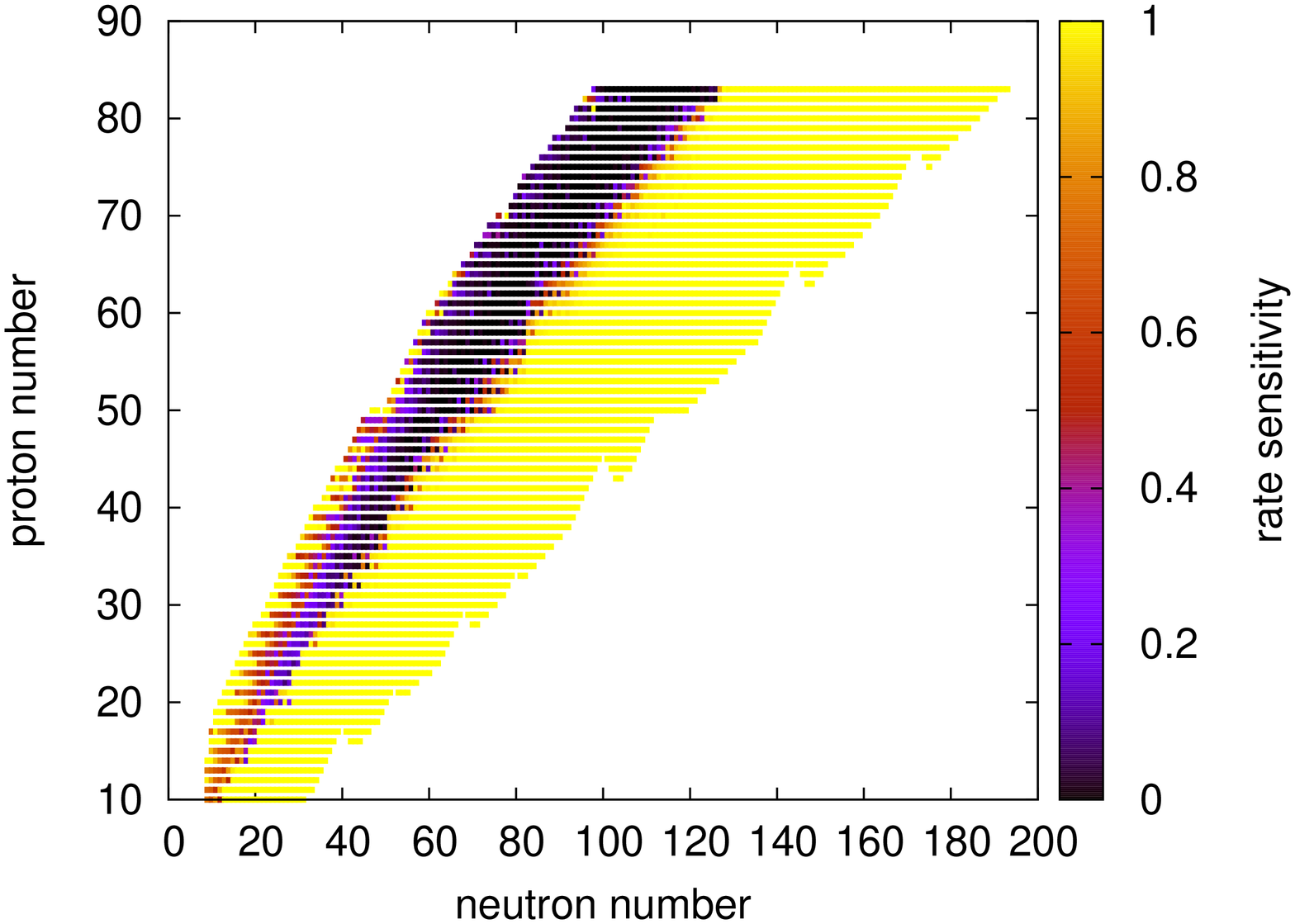}
\caption{Sensitivities of stellar proton capture rates at 1.5 GK to a variation of the $\gamma$-width. See the electronic edition of the Journal for a color version
of this figure.\label{fig:3Dpggamm}}
\end{figure}

\begin{figure}
%\plottwo{f2.eps}{f2_color.eps}
\includegraphics[angle=-90,width=\columnwidth]{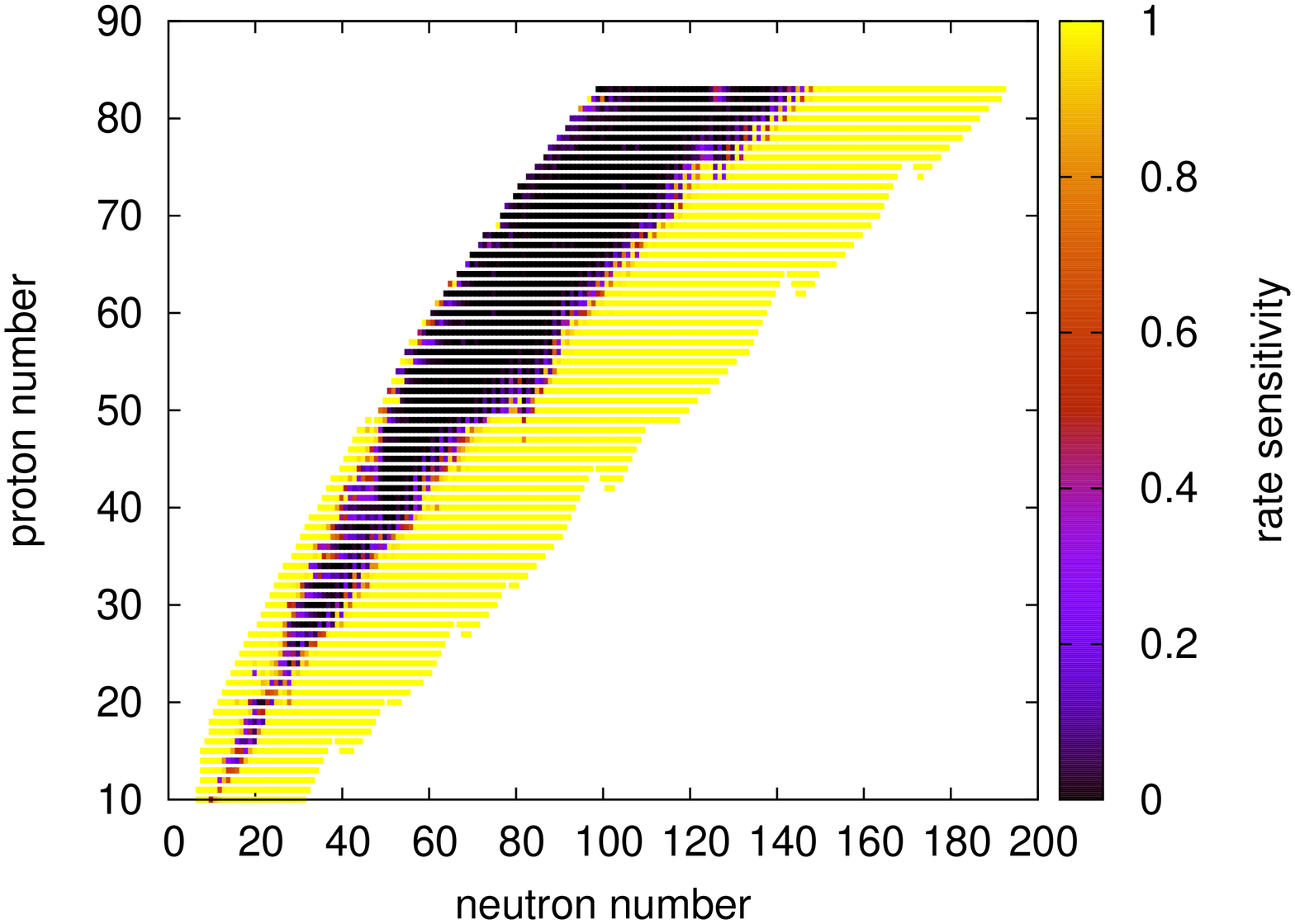}
\caption{Sensitivities of stellar $\alpha$ capture rates at 1.5 GK to a variation of the $\gamma$-width. See the electronic edition of the Journal for a color version
of this figure.\label{fig:3Daggamm}}
\end{figure}

\begin{figure}
%\plottwo{f2.eps}{f2_color.eps}
\includegraphics[angle=-90,width=\columnwidth]{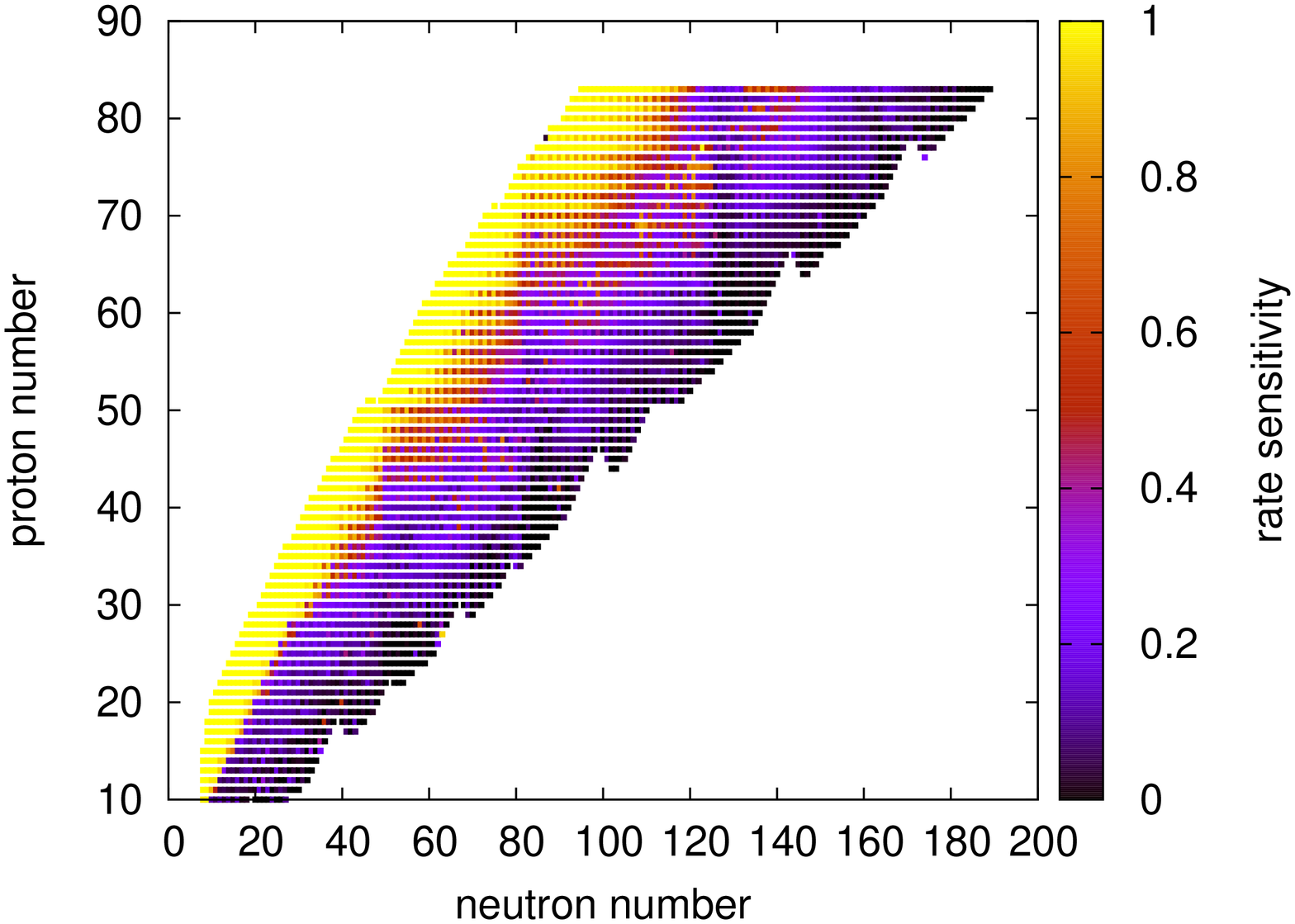}
\caption{Sensitivities of stellar neutron capture rates at 0.3 GK to a variation of the neutron width. See the electronic edition of the Journal for a color version
of this figure.\label{fig:3Dngneut}}
\end{figure}

\begin{figure}
%\plottwo{f2.eps}{f2_color.eps}
\includegraphics[angle=-90,width=\columnwidth]{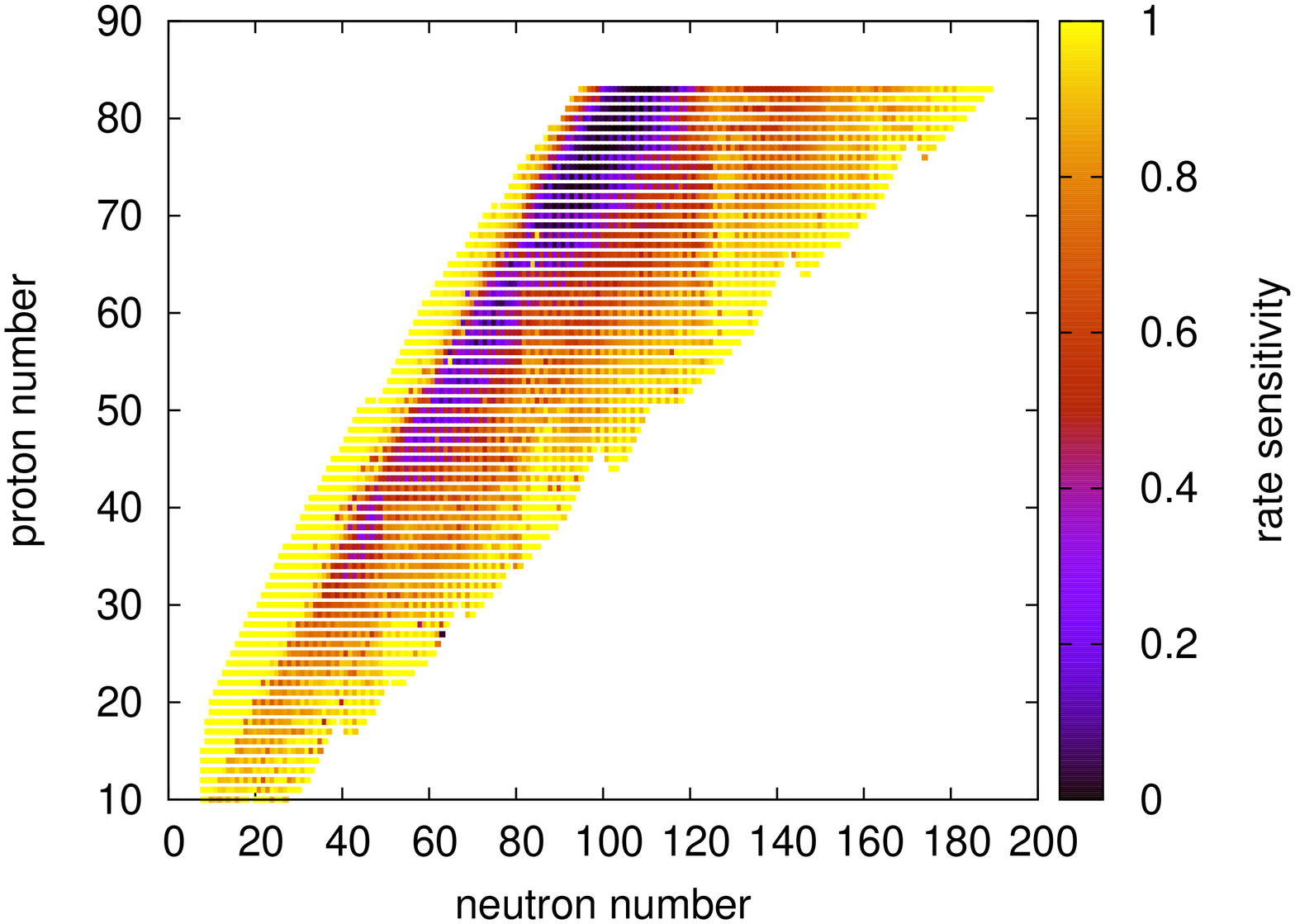}
\caption{Sensitivities of stellar neutron capture rates at 0.3 GK to a variation of the $\gamma$-width. See the electronic edition of the Journal for a color version
of this figure.\label{fig:3Dnggamm}}
\end{figure}

\begin{figure}
%\plottwo{f2.eps}{f2_color.eps}
\includegraphics[angle=-90,width=\columnwidth]{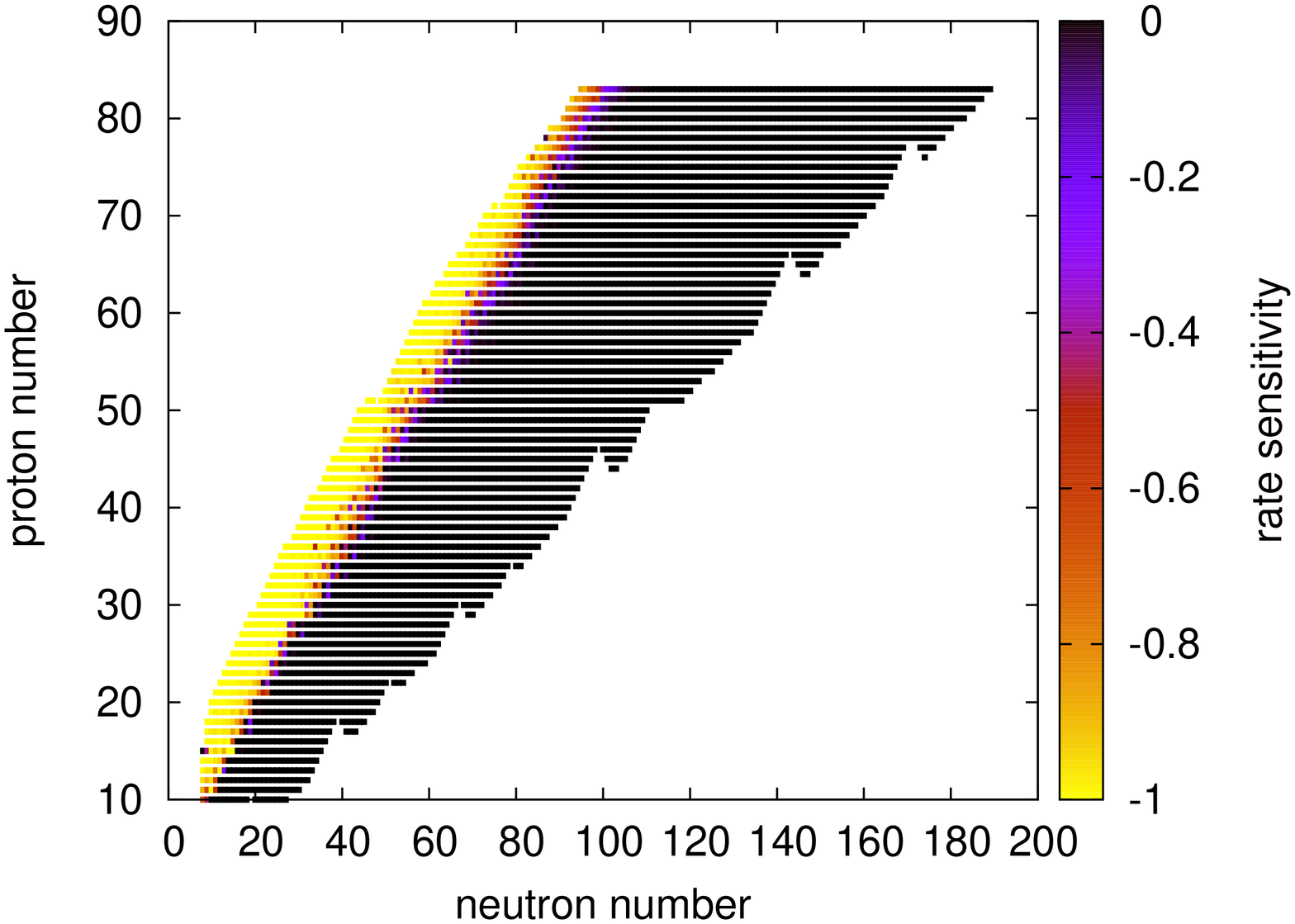}
\caption{Sensitivities of stellar neutron capture rates at 0.3 GK to a variation of the proton width. See the electronic edition of the Journal for a color version
of this figure.\label{fig:3Dngprot}}
\end{figure}

A more general view is provided in Figures \ref{fig:3Dpgprot}--\ref{fig:3Dngprot}, showing the values of sensitivities across the nuclear chart for a given reaction, width, and temperature. Sensitivities for proton capture rates at $T=1.5$ GK are shown in Figures \ref{fig:3Dpgprot} and \ref{fig:3Dpggamm}. As seen in Figure \ref{fig:3Dpgprot}, the (p,$\gamma$) rates are maximally sensitive to changes in the proton width all across the nuclear chart except for the lighter range of nuclei at the proton dripline. These have very low reaction $Q$-values and the statistical model is not applicable \citep{rtk97,raureview}. Their stellar rate sensitivity is dominated by the sensitivity to the $\gamma$-width, as depicted in Figure \ref{fig:3Dpggamm}. For proton-rich nuclei with small to moderate capture $Q$-values (up to a few MeV), the sensitivities to proton or $\gamma$-widths are complementary, as a comparison of the two figures reveals. For proton captures on neutron-rich nuclei, the rates are sensitive to both widths because the total width in the denominator of $\mathcal{F}$ in Equation (\ref{eq:f}) is dominated by the neutron width. This is case 2 discussed above. Therefore they are also sensitive to, and change inversely proportionally to, a variation of the neutron widths in that region of the nuclear chart.

A somewhat similar behavior is found for ($\alpha$,$\gamma$) rates. Due to the higher Coulomb barrier, and thus overall smaller width, the rates are sensitive to variations of the $\alpha$-width all over the nuclear chart. Therefore we only show the sensitivities to changes in the $\gamma$-width in Figure \ref{fig:3Daggamm}. Similar to proton captures, $\alpha$-capture rates on neutron-rich nuclei are sensitive to $\alpha$-, $\gamma$-, and neutron widths, with the neutron width dominating the denominator and thus leading to an inverse proportional dependence on variations of the neutron width.

The sensitivities of neutron capture reactions at 0.3 GK (close to typical s-process temperatures) to variations of the neutron- and $\gamma$-widths are shown in Figures \ref{fig:3Dngneut} and \ref{fig:3Dnggamm}, respectively. Again, the sensitivities are basically complementary, with decreasing sensitivities to the neutron width with decreasing $Q$-value towards the neutron-rich side of stability. The exception are (n,$\gamma$) rates involving target nuclei close to the proton dripline. These are sensitive to both, neutron and $\gamma$-widths, in the same manner and inversely sensitive to the proton width which determines the denominator of $\mathcal{F}$, as can be seen in Figure \ref{fig:3Dngprot}.

\subsubsection{Cross sections}
\label{sec:cstables}

Although the laboratory cross section, i.e.\ the cross section of a reaction proceeding on a target nucleus in the g.s.\ as defined in \S~\ref{sec:ratedef}, cannot be used to calculate the stellar reaction rate unless $X=1$, experiments may be performed to better constrain input required for the prediction of the rate. This is only possible when the measured cross section is indeed sensitive to the nuclear property in question. For this purpose, tables of laboratory cross section sensitivities to the same variation of the same quantities as for the rates shown in \S~\ref{sec:ratetables} are provided here. Table \ref{tab:xstable} is structured similarly to Table \ref{tab:sensi} but it contains the sensitivities to the four widths at a varying number of projectile energies and no distinction is made regarding the sign of the reaction $Q$ value. Also, the g.s.\ contribution $X$ is not given, as it does not apply here. Table \ref{tab:xstable} is organized as follows: The first column contains the abbreviated element name of the target nucleus $A$ and the next two columns give its charge $Z_A$ and mass number $M_A$. The fourth and fifth columns identify the reaction by specifying projectile $a$ and ejectile $b$, respectively. The sixth column gives the number of energies $n_\mathrm{En}$ for which values are present. This is followed by 50 blocks of 5 columns each. In each block the first entry is the center of mass projectile energy $E$ in MeV and the next four entries are the sensitivities to a variation of the $\gamma$-, neutron-, proton-, and $\alpha$-width ($^\mathrm{cs} s_\mathrm{g}$, $^\mathrm{cs} s_\mathrm{n}$, $^\mathrm{cs} s_\mathrm{p}$, $^\mathrm{cs} s_\mathrm{a}$, in that order). Empty blocks, i.e.\ blocks beyond $n_\mathrm{En}$, contain an energy of $-1.0$ and all sensitivities are zero. Empty blocks appear when the number of valid energy entries is smaller than 50. Valid energies are those below 14 MeV for which at least one of the sensitivities is non-zero. If there is a sequence of energies with only zero sensitivity, e.g.\ for endothermic reactions, only the entries for the highest of these energies is shown to indicate that all sensitivities are zero below this energy. Reactions with fewer than five valid energies are omitted. The energy cut-off at 14 MeV still encompasses all astrophysically relevant energies and at the same time ensures that the reaction model used is sufficient.

The cases discussed in \S~\ref{sec:ratetables} regarding the dependence of rates on the widths appearing in the factor $\mathcal{F}$ also apply to cross sections. The energy dependence of the sensitivities, however, are stronger than the dependence on the temperature in the rates. The energies shown in Table \ref{tab:xstable} cover a larger range than required for the integration of the reaction rates. Moreover, the energy dependence in the cross sections is not diluted by an integration as for the rates. In particular, large changes in the sensitivities appear at energies where a new reaction channel opens. This underlines the fact that it is necessary to experimentally determine the cross sections as close as possible to the astrophysically relevant energy range because the cross sections are most likely depending on different widths at higher energies. Again, this can be best seen when examining charged-particle capture. At high energies, the energy dependence of the cross section and its uncertainty is governed by the $\gamma$-width whereas in the astrophysical energy range it is determined by the charged-particle width.

\begin{figure}
%\plottwo{f2.eps}{f2_color.eps}
\includegraphics[angle=-90,width=\columnwidth]{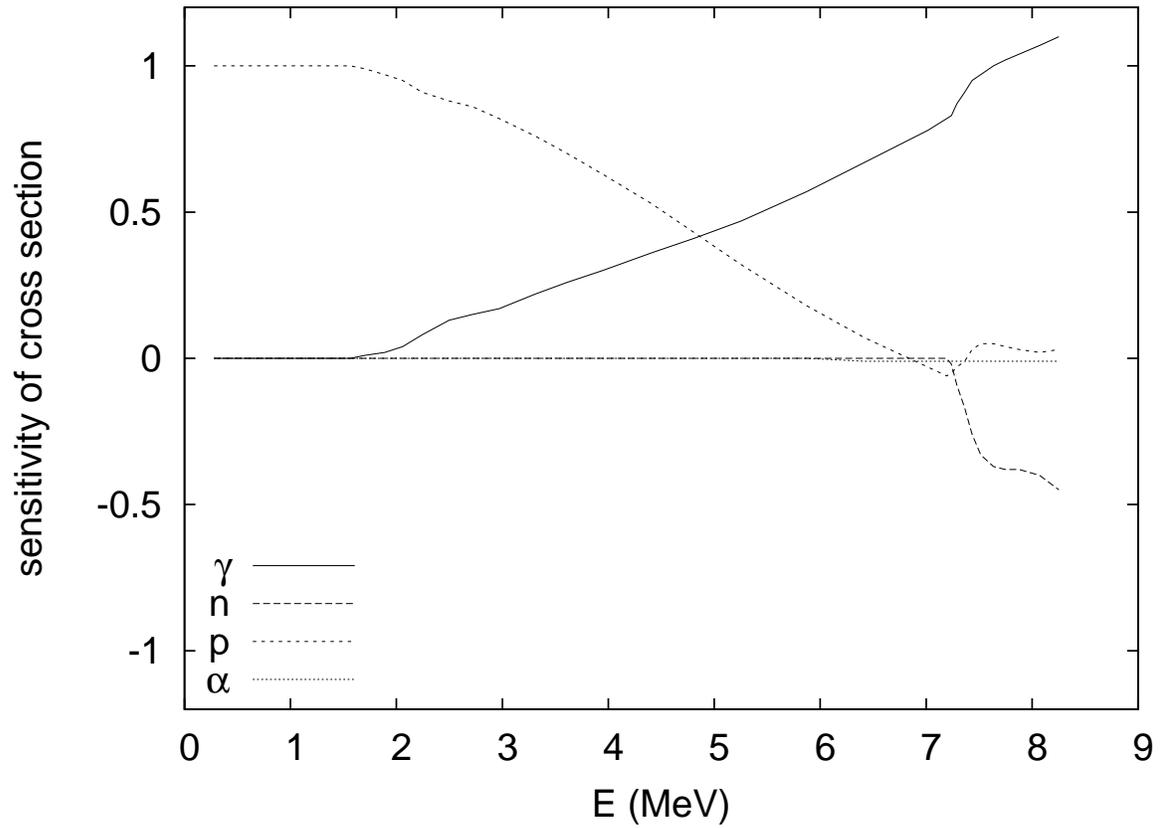}
\caption{Sensitivity of the laboratory cross section of $^{96}$Ru(p,$\gamma$)$^{97}$Rh as function of c.m.\ energy. The astrophysically relevant energy range is $1.5\leq E\leq 3.85$ MeV (for $2\leq T \leq 3$ GK).\label{fig:ruplot}}
\end{figure}

For illustration, Figure \ref{fig:ruplot} shows the sensitivities of the laboratory reaction cross section of $^{96}$Ru(p,$\gamma$)$^{97}$Rh as function of c.m.\ energy. This reaction appears in p-process studies \citep{arngorp,reifarthGSI} and the astrophysically relevant energy range is $1.5\leq E\leq 3.85$ MeV (for plasma temperatures $2\leq T \leq 3$ GK). It can be clearly seen that the impact of each width type drastically changes from lower to higher energy and thus also the relevance of each type of width strongly depends on energy. At low energy, the cross section is only sensitive to the proton width and all uncertainties therein fully affect its prediction. At the high end of the shown energy range, the impact of the proton width is negligible and the by far highest sensitivity is to the $\gamma$-width. Uncertainties in the $\alpha$-width are irrelevant for this reaction over the whole energy range, whereas the importance of the neutron width quickly increases above the (p,n) threshold with increasing energy. Since the neutron width appears in the denominator of $\mathcal{F}$ in Equation (\ref{eq:f}), the size of the cross section varies inverse proportionally to the size of the neutron width. If a comparison between experimental data and theory finds a discrepancy above, say, 7 MeV, this is still inconsequential for the astrophysical rate. On the other hand, if theory reproduced the cross sections above 7 MeV well, the astrophysical rate would still be unconstrained because it is sensitive to a different nuclear property.

\begin{figure}
%\plottwo{f2.eps}{f2_color.eps}
\includegraphics[angle=-90,width=\columnwidth]{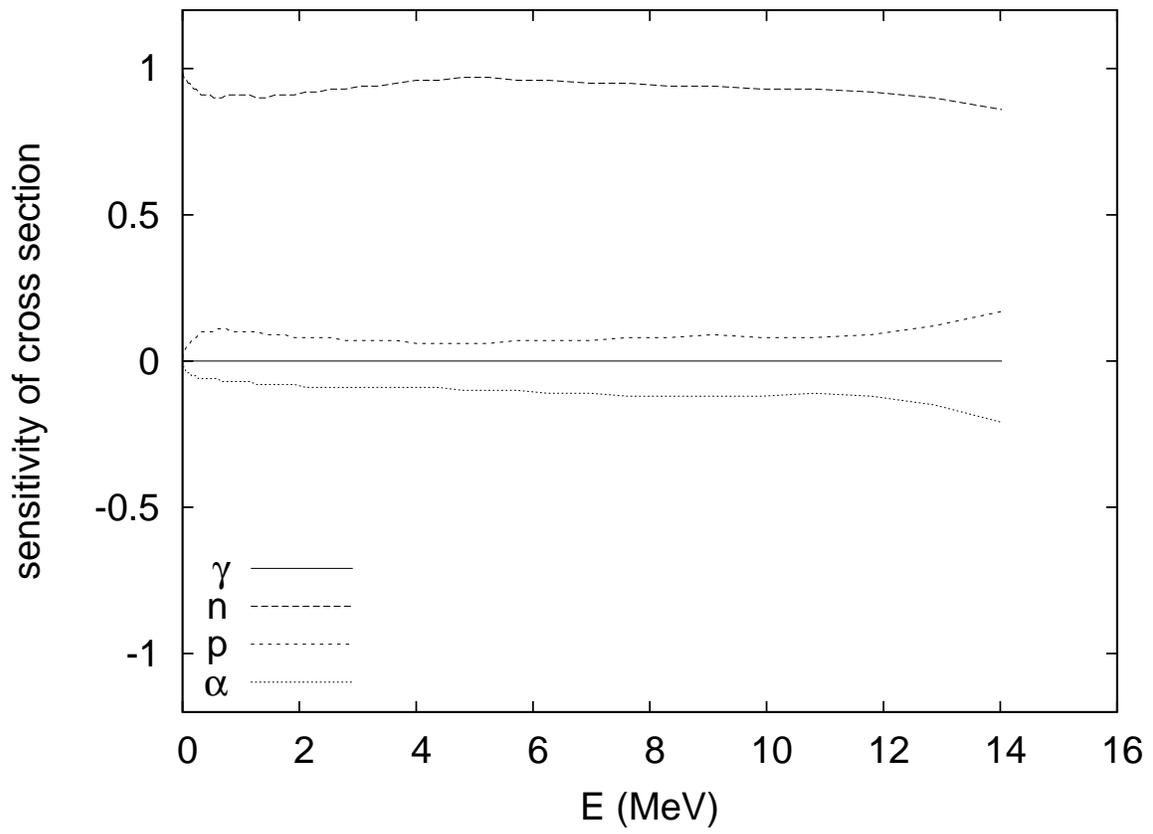}
\caption{Sensitivity of the laboratory cross section of $^{64}$Ge(n,p)$^{64}$Ga as function of c.m.\ energy.\label{fig:ge64np}}
\end{figure}

Not all laboratory cross sections have strongly varying sensitivities with energy. Figure \ref{fig:ge64np} shows the situation for $^{64}$Ge(n,p)$^{64}$Ga, important in $\nu$p-process nucleosynthesis \citep{froh,pruet}. It is exclusively sensitive to the neutron width across the shown range of energies.

\begin{figure}
%\plottwo{f2.eps}{f2_color.eps}
\includegraphics[angle=-90,width=\columnwidth]{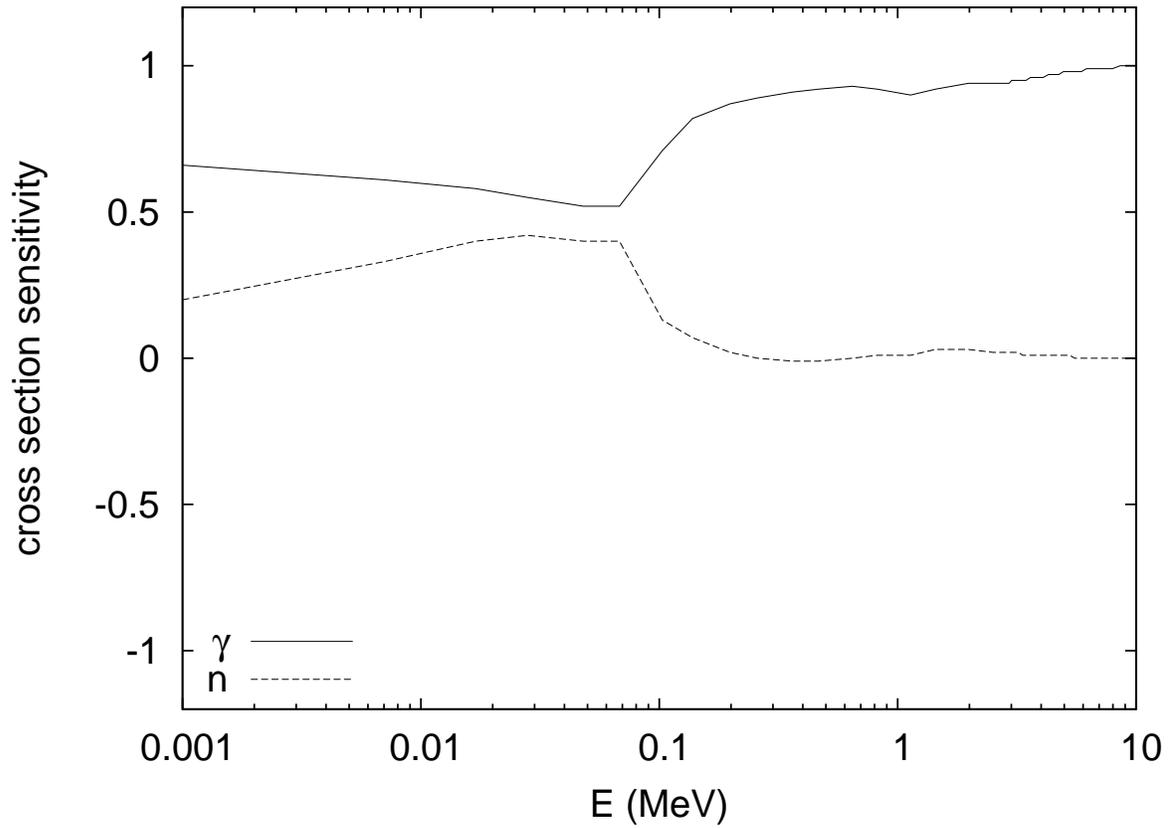}
\caption{Sensitivity of the laboratory cross section of $^{180}$Hf(n,$\gamma$)$^{181}$Hf as function of c.m.\ energy; note the logarithmic energy scale. The astrophysically relevant energy range is $0.\leq E\leq 0.1$ MeV for s-process nucleosynthesis.\label{fig:hf180ng}}
\end{figure}

\begin{figure}
%\plottwo{f2.eps}{f2_color.eps}
\includegraphics[angle=-90,width=\columnwidth]{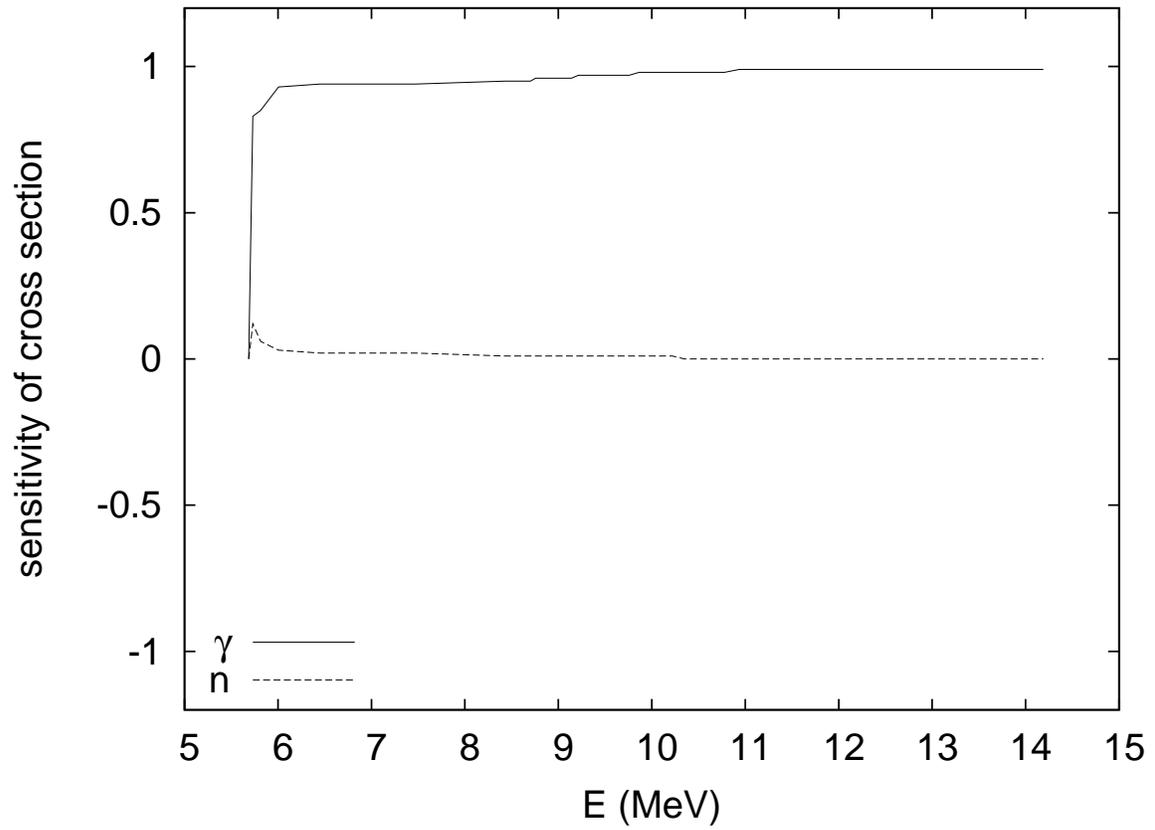}
\caption{Sensitivity of the laboratory cross section of $^{181}$Hf($\gamma$,n)$^{180}$Hf as function of c.m.\ energy. The sensitivities to variations of proton or $\alpha$ width are zero and not shown here.\label{fig:hf181gn}}
\end{figure}

As another example, the cross section of $^{180}$Hf(n,$\gamma$)$^{181}$Hf seems to be mostly sensitive to the $\gamma$-width across the considered energy range. A closer inspection, however, shows that there is a quite similar sensitivity to both $\gamma$- and neutron width at typical energies of $0-100$ keV encountered in the s-process \citep{sprocess}. This is shown in Figure \ref{fig:hf180ng}. The \textit{laboratory} cross section of the reverse reaction $^{181}$Hf($\gamma$,n)$^{180}$Hf, on the other hand, exhibits only sensitivity to the $\gamma$-width, as can be seen in Figure \ref{fig:hf181gn}. Although there is reciprocity between single transitions according to the reciprocity theorem of nuclear reactions \citep{BW52}, laboratory cross sections do not obey reciprocity because different transitions are included in the forward and reverse reaction, as can be easily inferred from Figure \ref{fig:transitions} and also from the definition of the laboratory cross section provided in \S~\ref{sec:ratedef}.

\subsection{Dependence of widths on input quantities}
\label{sec:dependence}

The tables in \S~\ref{sec:sensitables} allow to judge the importance of a width for the stellar reaction rate or the laboratory cross section. It is then important to consider which nuclear properties determine a width and its uncertainty. We do not want to repeat the detailed discussion in \citet{raureview} here but rather give a short summary of what enters the determination of a width. It equally applies to the widths of individual resonances to be used with Equation (\ref{eq:BWrate}) and to the averaged widths appearing in Equation (\ref{eq:HFrate}).

The transition widths $W^\lambda$ in Equation (\ref{eq:f}) are actually derived by accounting for all energetically possible transitions from a compound level $(E_C,J_C,\pi_C)$ to levels in channel $\lambda$ \citep[see, e.g.,][and references therein]{hwfz,ctt},
\begin{eqnarray}
W^\lambda_{E_C,J_C,\pi_C}&=&\left( \sum_{n=0}^{n_\mathrm{last}^\lambda} W^{(E_C,J_C,\pi_C)\rightarrow (E_n^\lambda,J_n^\lambda,\pi_n^\lambda)} \right) \nonumber \\
&& + \left( \int_{E_\mathrm{last}^\lambda}^{E_C-S_\mathrm{proj}} \sum_{J^\lambda,\pi^\lambda} W^{(E_C,J_C,\pi_C)\rightarrow (E^\lambda,J^\lambda,\pi^\lambda)} \rho^\lambda(E^\lambda,J^\lambda,\pi^\lambda)\,dE^\lambda \right) \quad.
\label{eq:transcoeff}
\end{eqnarray}
Optical potentials are required to compute the particle transition probabilities by solving the (time-independent) Schr\"odinger equation. Photon strength functions, on the other hand, determine the strengths of the $\gamma$-transitions. In addition, the calculated widths depend on the discrete level scheme included up to the $n_\mathrm{last}^\lambda$-th level at energy $E_\mathrm{last}^\lambda$. They may also depend on the NLD $\rho^\lambda$ when transitions to states above $E_\mathrm{last}^\lambda$ are possible (see Figure \ref{fig:transitions}).

In principle, the NLD enters in two ways. Firstly, the NLD at the compound formation energy determines which reaction model should be used (see \S~\ref{sec:applicability}). Once the model has been chosen, however, it does not enter the calculation anymore (unless using a modified, non-standard model, see, e.g., \citet{raureview}). As discussed in \S~\ref{sec:applicability}, the choice cannot be made from within the model and therefore comprises a type 1 uncertainty which cannot be estimated systematically as explained in \S~\ref{sec:intro}. Secondly, the NLD below the compound formation energy may be required to calculate the widths $W$ appearing in Equations (\ref{eq:BWrate}) and (\ref{eq:HFrate}) when transitions to unresolved or unknown excited states are important, as shown in Equation (\ref{eq:transcoeff}).
Because NLD studies often take their motivation from an application to astrophysics, the sensitivities of reaction rates and cross sections to variation of the NLD is studied in detail in the following \S~\ref{sec:nldvari}.

Nuclear masses are important insofar as mass differences determine $Q$ values and separation energies and thus also the relative energies of the possible transitions, as shown in Figure \ref{fig:transitions}. This then impacts the calculation of the transition widths. Therefore, uncertainties in masses become relevant when they lead to sizeable changes in separation energies, exceeding a few tens of percent. Close to stability the mass differences are known with such high accuracy that essentially no uncertainty is introduced into the rates from masses. This holds as long as separation energies are large. Approaching the driplines, separation energies become smaller and uncertainties in masses larger. Nevertheless, whenever experimental information is available, the uncertainties are negligible. For most nuclei along the dripline, only theoretical mass predictions are available and different mass models yield largely different values. It should be noted, however, that mass differences are still thought to be predicted more accurately than masses of single nuclei.

Sensitivities of rates to a variation of masses are not shown here for three reasons, the first reason being the mentioned fact that the separation energies for the majority of reactions in the nuclear chart are known with sufficient accuracy. The second reason is that masses cannot be varied independently as they simultaneously affect the separation energies appearing in several reactions. The third reason is that for reactions with low separation energies and large mass uncertainties the ratio between forward and reverse rate is more affected by the mass uncertainties than the forward rate itself. This can be seen in Equations (\ref{eq:recipart}) and (\ref{eq:reciphoto}), where the exponential dependence on the reaction $Q$-value is explicitly shown. In most cases, the change in the balance between forward and reverse rate will affect astrophysical applications more than the pure change in the reactivity. This is only different very close to the driplines, when a change in the masses leads to a largely different range of relative channel energies and consequently of accessible levels. Then a recalculation of the reactivity is called for.

\subsection{Sensitivity tables for NLD variations}
\label{sec:nldvari}

\subsubsection{Rates}
\label{sec:nldrates}

Table \ref{tab:nldsensi} shows the sensitivities of the astrophysical reaction rates to variations in the NLD by a factor $v_q=2$. The table is constructed similarly to Table \ref{tab:sensi} but with several additional columns. The first column contains the abbreviated element name of the target nucleus $A$ and the next two columns give its charge $Z_A$ and mass number $M_A$. The fourth and fifth columns identify the reaction by specifying projectile $a$ and ejectile $b$, respectively. The sixth to tenth column gives the excitation energy $E_\mathrm{last}$ of the highest-lying, discrete excited level included in the calculation for the four reaction channels defined by emission of $\gamma$ (g), neutron (n), proton (p), or $\alpha$-particle (a). This is followed by 24 blocks for the same 24 temperatures as in Table \ref{tab:sensi}. Each block contains eight columns showing the sensitivity $^\mathrm{r} _\mathrm{D} s_q$ to a variation of the NLD in the $\gamma$-, neutron-, proton-, and $\alpha$-channel. The first set of four columns in a block show the sensitivities ($^\mathrm{r} _\mathrm{D} s_\mathrm{g}$, $^\mathrm{r} _\mathrm{D} s_\mathrm{n}$, $^\mathrm{r} _\mathrm{D} s_\mathrm{p}$, $^\mathrm{r} _\mathrm{D} s_\mathrm{a}$, in that order) when including experimentally known excited states in each channel up to the excitation energy $E_\mathrm{last}^\mathrm{g}$, $E_\mathrm{last}^\mathrm{n}$, $E_\mathrm{last}^\mathrm{p}$, and $E_\mathrm{last}^\mathrm{a}$, respectively. The second set of four columns in a block give sensitivities ($^\mathrm{r} _\mathrm{D} s_\mathrm{g}^0$, $^\mathrm{r} _\mathrm{D} s_\mathrm{n}^0$, $^\mathrm{r} _\mathrm{D} s_\mathrm{p}^0$, $^\mathrm{r} _\mathrm{D} s_\mathrm{a}^0$, in that order) calculated without excited states, i.e., using the NLD immediately above the ground state.

It should be noted that the shown sensitivities actually arise from a combination of two sensitivities, although they were directly inferred from the variation of the NLD and its impact on the final rate, following Equation (\ref{eq:sensi}). The sensitivity of a width to a change in the NLD is folded with the sensitivity of the rate to a change in the width. The latter sensitivity is discussed in \S~\ref{sec:ratetables} and shown separately in Table \ref{tab:sensi}. Obviously, when a rate is not sensitive to a change in a width, it will also not be modified when this width is changed by a variation of the NLD. As discussed in \S~\ref{sec:ratetables}, it should also be noted that a relevant width can appear in the numerator or denominator of the ratio $\mathcal{F}$ in Equation (\ref{eq:f}). If it is in the denominator, the rate will change inversely proportional to the change in the width, i.e., it will become smaller when the width is increased. As before, this is identified by a negative sign of the sensitivity given in the tables here.

\begin{figure}
%\plottwo{f2.eps}{f2_color.eps}
\includegraphics[angle=-90,width=\columnwidth]{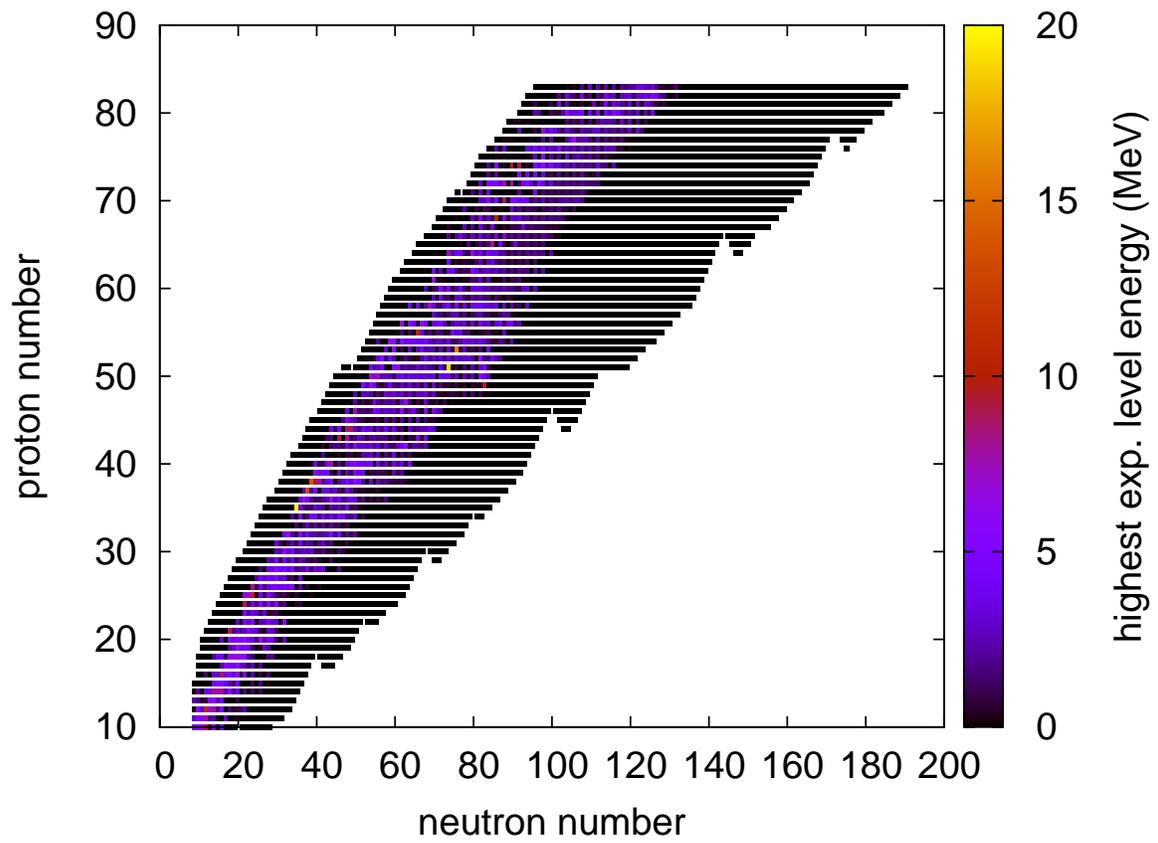}
\caption{Energy $E_\mathrm{last}$ of the highest included discrete, experimental level; above that energy, a theoretical NLD is used. See the electronic edition of the Journal for a color version
of this figure.\label{fig:maxlevel}}
\end{figure}

The standard approach in the calculation of reaction rates for astrophysics in a Hauser-Feshbach statistical model is to include a number of discrete excited states, when experimentally known. The discrete levels used are a selection from \citet{ENSDF}, including levels with well determined spins and parities and up to the energy at which the level scheme can still be considered complete. A predicted NLD is only used above the energy of the last included experimental level $E_\mathrm{last}$.  Figure \ref{fig:maxlevel} shows $E_\mathrm{last}$ for all considered nuclei. This reduces some of the rate uncertainty but obviously only for nuclei with accurately known level schemes. Moreover, this procedure can only reduce the uncertainty in the channels where transitions to the included states dominate the rate. This occurs mostly in the particle channels but is not the case for the $\gamma$-channel of a reaction due to the higher $Q$-value and thus larger range of possible relative energies. It has been shown in \citet{raugammaenergies} that the dominant $\gamma$-transitions in astrophysical capture lead to levels $2-4$ MeV below the formation energy of the compound nucleus. Even close to stability, complete level schemes are not known at this excitation energy. In captures with low $Q$-value, on the other hand, transitions to the lowest excited states are favored and the knowledge of a few levels above the compound g.s.\ will be sufficient. To see the impact of the included levels, the sensitivities $^\mathrm{r} _\mathrm{D} s_q$ and $^\mathrm{r} _\mathrm{D} s_q^0$, with and without experimental levels, respectively, can be compared for a specific reaction rate at a given temperature. The difference will be larger, the larger $E+Q-E_\mathrm{last}$ is in the reaction channel, for channels with $E_\mathrm{last}>0$ and $Q$ being the reaction $Q$-value applicable to the channel. The energy $E$ mostly contributing to the reaction rate at a given temperature is the location of the maximum contribution within the relevant energy window and can be found in \citet{energywindows}.

\begin{figure}
%\plottwo{f2.eps}{f2_color.eps}
\includegraphics[angle=-90,width=\columnwidth]{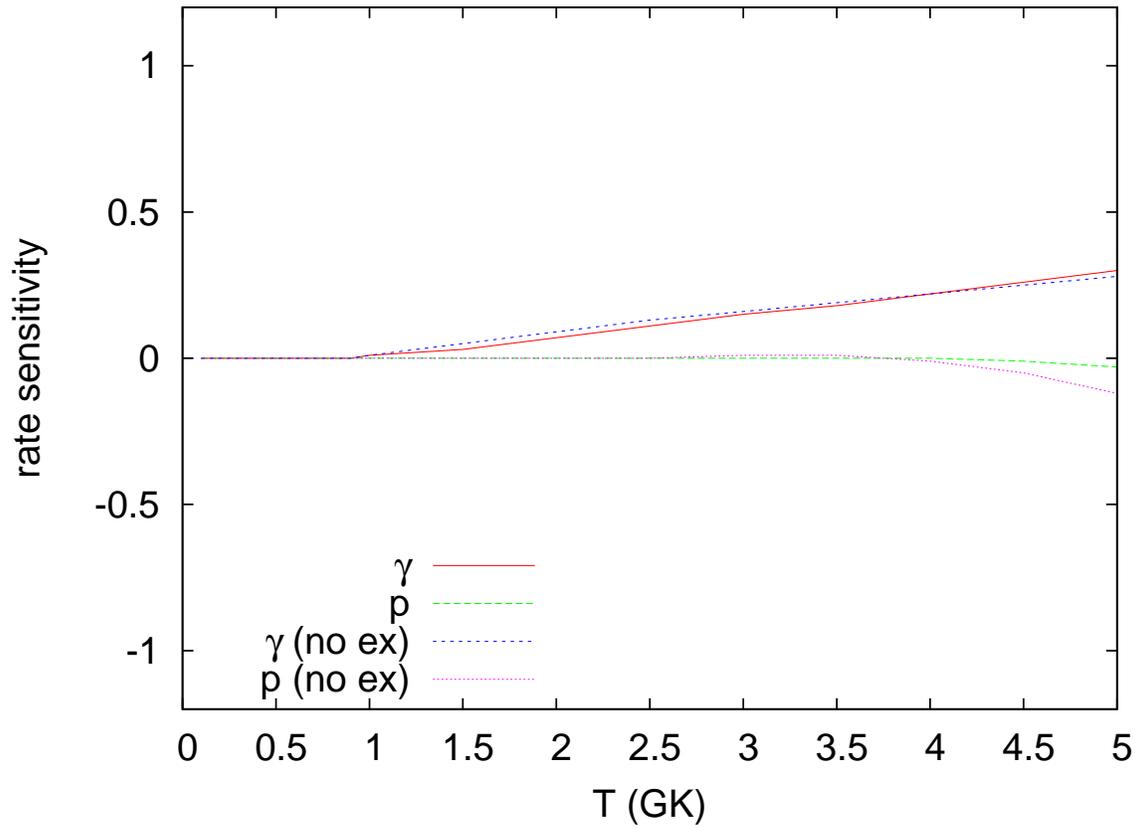}
\caption{Sensitivity of the stellar rate of $^{96}$Ru(p,$\gamma$)$^{97}$Rh to a variation of the NLD (with and without inclusion of experimental excited states) in the proton and $\gamma$ channel as function of plasma temperature. See the electronic edition of the Journal for a color version of this figure.\label{fig:ru96nldrate}}
\end{figure}

As an example, Figure \ref{fig:ru96nldrate} shows the stellar rate sensitivities of $^{96}$Ru(p,$\gamma$)$^{97}$Rh to a variation of the NLD in the proton- and $\gamma$-channel, with and without inclusion of discrete excited states. This is to be compared to Figure \ref{fig:ru96rate}, showing the sensitivity of this rate to width variations. Firstly, it can be seen that the inclusion of excited states does not change much in the NLD sensitivity here. Secondly, the rate is not sensitive to the NLD variation in the proton channel although it shows the highest sensitivity to the proton width. The sensitivity to a NLD variation in the $\gamma$-channel, on the other hand, is the same as the overall sensitivity to a variation of the $\gamma$-width. This can be understood by examining which transitions are important. As pointed out above, the most important $\gamma$-transitions from the compound state end at levels with high excitation energy and thus the NLD directly determines the size of the $\gamma$-width. In the proton channel, the most important transitions go to low-lying states or, as in this case, to the g.s. This fact is confirmed by a value of the g.s.\ contribution to the stellar rate $X\approx 1$.

Similar considerations apply for the rates of $^{144}$Sm($\alpha$,$\gamma$)$^{148}$Gd and $^{132}$Sn(n,$\gamma$)$^{133}$Sn. Their sensitivities to width variations were discussed in \S~\ref{sec:ratetables} and shown in Figures \ref{fig:rate_sm144ag} and \ref{fig:rate_sn132ng}. Both rates are found to be insensitive to variations of the NLD. The rate of $^{144}$Sm($\alpha$,$\gamma$)$^{148}$Gd is only sensitive to the $\alpha$-width and has a strong g.s.\ contribution to the stellar rate. This implies that $\alpha$ transitions to the g.s.\ of $^{144}$Sm are dominating which makes the width insensitive to a variation of the NLD at higher excitation energy. Although the rate of $^{132}$Sn(n,$\gamma$)$^{133}$Sn is almost exclusively sensitive to the $\gamma$-width, it shows also a very low sensitivity to any NLD variation. The dominating contribution to the $\gamma$-width stems from the g.s.\ $\gamma$-transition due to the low reaction $Q$-value.

In summary, a variation of the NLD only affects rates when the NLD is changed in a channel to which the rate is sensitive (see \S~\ref{sec:ratetables}) \textit{and} transitions to higher lying states are important. Therefore the rates are insensitive to NLD variations in the entrance channel of charged particle captures throughout the nuclear chart and only weakly sensitive in neutron captures. They are sensitive to a NLD variation in the $\gamma$ exit channel when they are sensitive to a variation of this $\gamma$-width at all. However, in these cases charged particle captures are also inverse proportionally sensitive to a variation of the neutron width, which enters the total width in the denominator of Equation (\ref{eq:f}), and therefore also to a NLD variation in this channel. Reactions with particles in all channels exhibit only a very weak sensitivity to NLD variations and this only close to stability for $Z_A\gtrsim 65$, if at all.

\subsubsection{Cross sections}
\label{sec:nldcs}

Similarly as for the cross section sensitivities to width variations, shown in \S~\ref{sec:cstables}, Table \ref{tab:nldxs} provides the sensitivities of laboratory cross sections when systematically varying the NLD. As for the rate sensitivities in Table \ref{tab:nldsensi}, these are shown with and without the inclusion of discrete excited states. These excited states were the same as those used in Table \ref{tab:nldsensi} and therefore the values for $E_\mathrm{last}$ remain the same and are not specified anymore. The structure of Table \ref{tab:nldxs} is close to that of Table \ref{tab:xstable}: The first column contains the abbreviated element name of the target nucleus $A$ and the next two columns give its charge $Z_A$ and mass number $M_A$. The fourth and fifth columns identify the reaction by specifying projectile $a$ and ejectile $b$, respectively. The sixth column gives the number of energies $n_\mathrm{En}$ for which values are present. This is followed by 50 blocks of 9 columns each. In each block the first entry is the center of mass projectile energy $E$ in MeV and the remaining set of eight columns contains the sensitivities to a variation of the NLD in the $\gamma$-, neutron-, proton-, and $\alpha$-channel. The first four columns in this set show the sensitivities ($^\mathrm{cs} _\mathrm{D} s_\mathrm{g}$, $^\mathrm{cs} _\mathrm{D} s_\mathrm{n}$, $^\mathrm{cs} _\mathrm{D} s_\mathrm{p}$, $^\mathrm{cs} _\mathrm{D} s_\mathrm{a}$, in that order) when including experimentally known excited states in each channel. The next four columns give sensitivities ($^\mathrm{cs} _\mathrm{D} s_\mathrm{g}^0$, $^\mathrm{cs} _\mathrm{D} s_\mathrm{n}^0$, $^\mathrm{cs} _\mathrm{D} s_\mathrm{p}^0$, $^\mathrm{cs} _\mathrm{D} s_\mathrm{a}^0$, in that order) calculated without excited states, i.e., using the NLD immediately above the ground state. As for Table \ref{tab:xstable}, empty blocks contain an energy of $-1.0$ and all sensitivities are zero. Photodisintegration and other reactions with negative reaction $Q$ value are included.

The sensitivites of laboratory cross sections to NLD variations display similar behavior as outlined above for the rates. In addition, it has to be remembered that the entrance width of a reaction appearing in the numerator of Equation (\ref{eq:f}) only contains the g.s.\ transition and therefore is not sensitive to the NLD at all. The total width in the denominator, however, contains the full width of the entrance channel, including possible contributions from the NLD as shown in Equation (\ref{eq:transcoeff}).

\begin{figure}
%\plottwo{f2.eps}{f2_color.eps}
\includegraphics[angle=-90,width=\columnwidth]{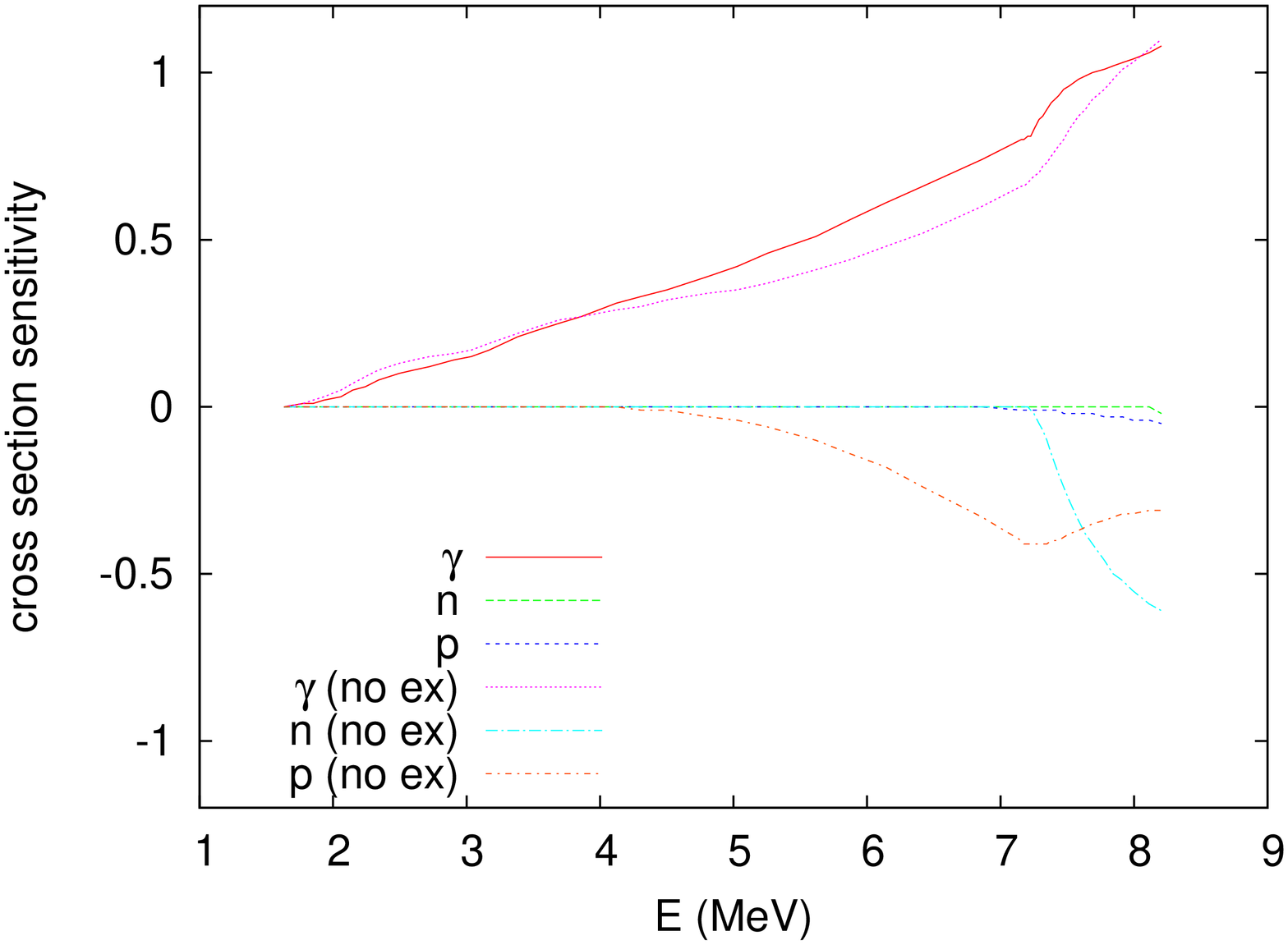}
\caption{Sensitivity of the laboratory cross section of $^{96}$Ru(p,$\gamma$)$^{97}$Rh to a variation of the NLD (with and without inclusion of experimental excited states) in the shown channels. See the electronic edition of the Journal for a color version of this figure.\label{fig:ru96nldxs}}
\end{figure}

As a specific example, Figure \ref{fig:ru96nldxs} shows the sensitivities of the laboratory cross section of $^{96}$Ru(p,$\gamma$)$^{97}$Rh to a variation of the NLD in the various channels, with and without inclusion of experimental excited states. The variation in the $\alpha$+$^{97}$Tc channel is not shown because the cross section is insensitive to it. The inclusion of excited states mostly affects the neutron channel above the threshold and it also moderately affects the proton channel above 5 MeV. When using the discrete excited states, there is no sensitivity to a change in the NLD in both channels. This implies that a sufficient range of excitation energies is covered by the experimental spectroscopic information. The sensitivity in the $\gamma$ channel is only weakly modified. Comparing to Figure \ref{fig:ruplot} it is apparent that transitions to excited states do not play a significant role in the proton channel below 5 MeV. The NLD variation does not result in a change of the cross section at those energies although the cross section would be very sensitive to changes of the proton width. Interestingly, above 5 MeV the cross section variation goes the opposite way as the NLD variation. This is due to the fact that the entrance width in the numerator of Equation (\ref{eq:f}) is independent of any NLD but the full entrance width appearing in the denominator depends on it to a certain fraction. Obviously, such a behavior is never found in stellar rates. Finally, the sensitivity in the $\gamma$-channel to the NLD variation is similar to the one of the cross section to the variation of the $\gamma$-width, showing that $\gamma$-transitions to higher lying excited states in $^{97}$Rh are dominating this width.

\begin{figure}
%\plottwo{f2.eps}{f2_color.eps}
\includegraphics[angle=-90,width=\columnwidth]{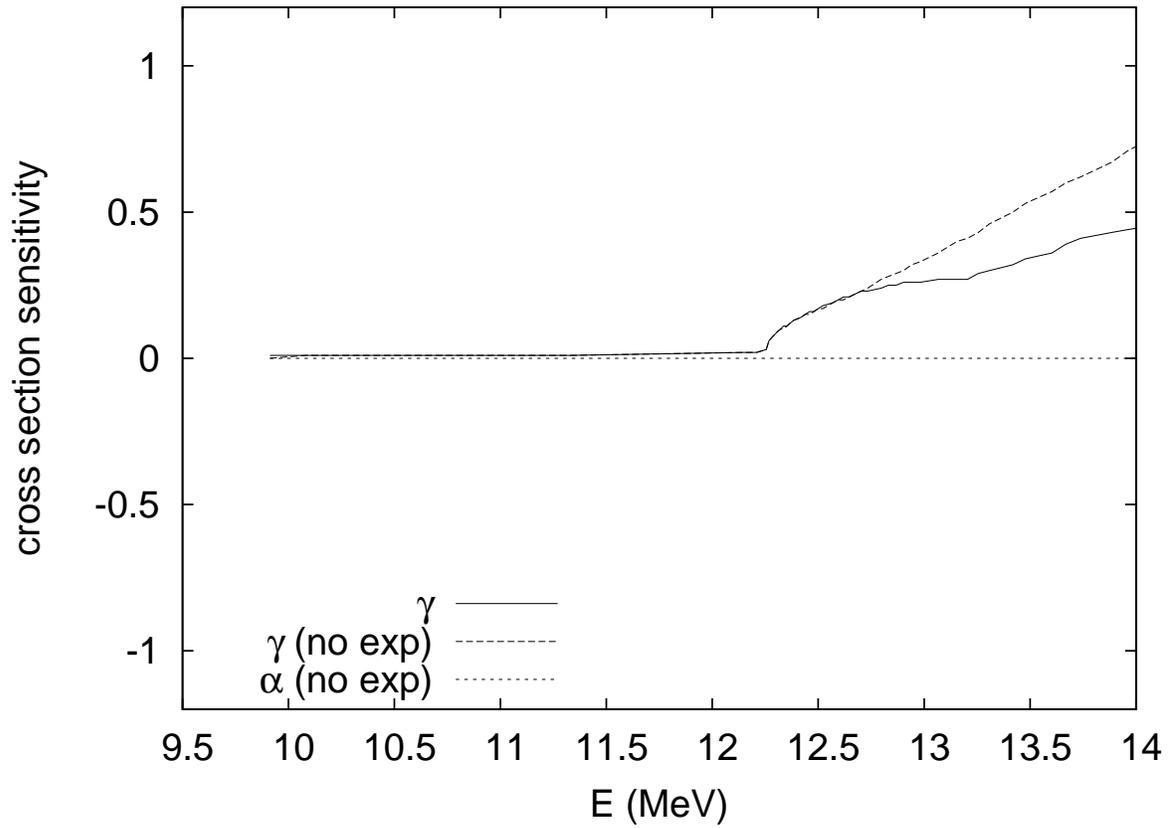}
\caption{Sensitivity of the laboratory cross section of $^{144}$Sm($\alpha$,$\gamma$)$^{148}$Gd to a variation of the NLD (with and without inclusion of experimental excited states) in the shown channels. \label{fig:sm144nldxs}}
\end{figure}

\begin{figure}
%\plottwo{f2.eps}{f2_color.eps}
\includegraphics[angle=-90,width=\columnwidth]{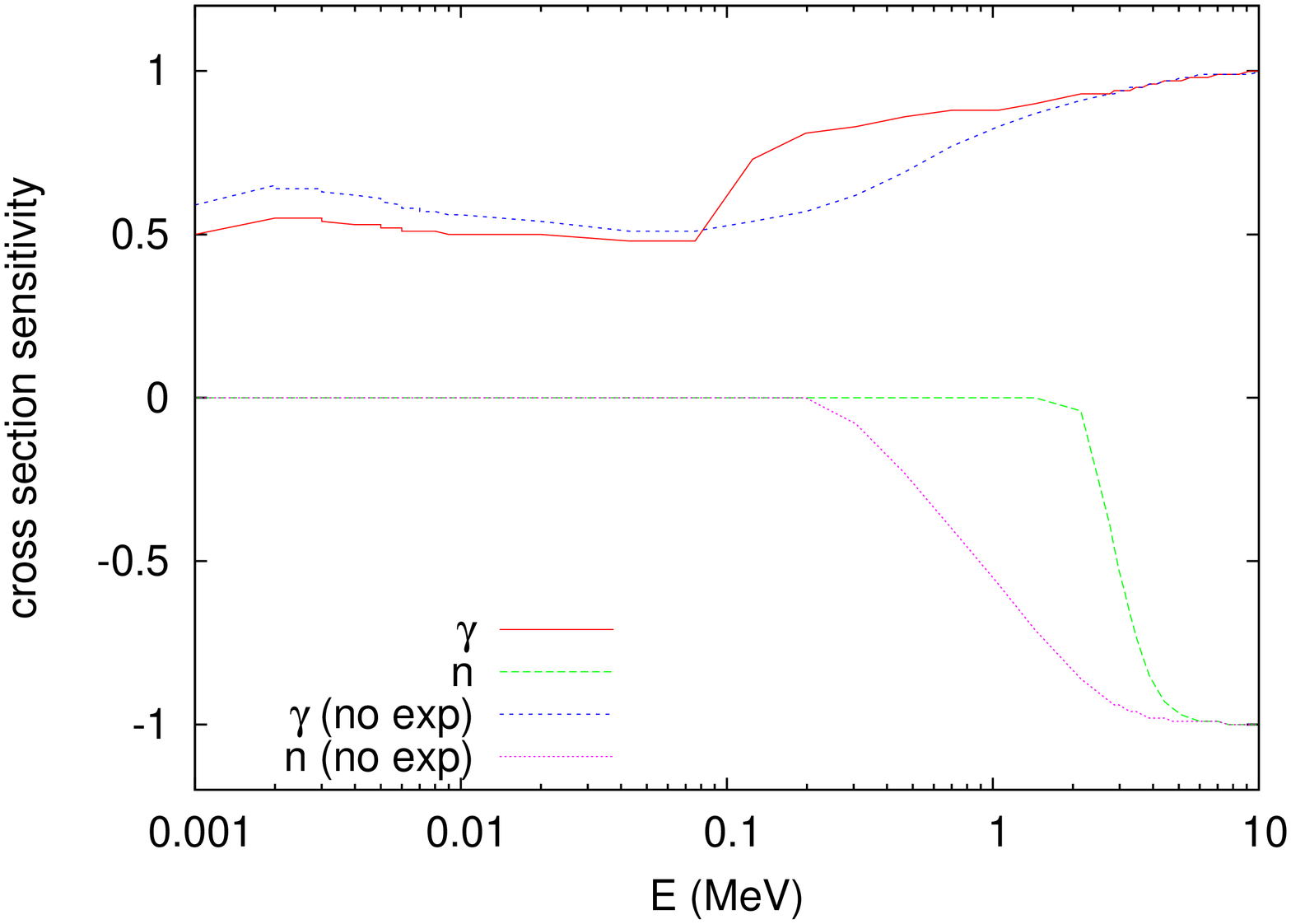}
\caption{Sensitivity of the laboratory cross section of $^{180}$Hf(n,$\gamma$)$^{181}$Hf to a variation of the NLD (with and without inclusion of experimental excited states) in the shown channels. See the electronic edition of the Journal for a color version of this figure.\label{fig:hf180nldxs}}
\end{figure}

As for the further examples given in the preceding sections, neither stellar rate nor laboratory cross section of $^{64}$Ge(n,p)$^{64}$Ga are sensitive to NLD variations. The laboratory cross section of $^{144}$Sm($\alpha$,$\gamma$)$^{148}$Gd above 12.3 MeV is weakly sensitive to the NLD in the $\gamma$+$^{148}$Gd channel but insensitive to a NLD variation in the $\alpha$+$^{144}$Sm channel although this is dominating the cross section (and rate) behavior (Figure \ref{fig:sm144nldxs}). Like before, the reaction $^{180}$Hf(n,$\gamma$)$^{181}$Hf exhibits a more complicated behavior at low energy. Comparing Figures \ref{fig:hf180nldxs} and \ref{fig:hf180ng} it is found that a NLD variation in the $\gamma$-channel leads to the same variation as the variation of the $\gamma$-width above 40 keV. Below this energy, the sensitivity to the NLD is weaker, indicating that $\gamma$-transitions to discrete states are of some relevance. Interestingly, the dependence on the NLD in the neutron channel is quite different from the sensitivity to the neutron width. While the cross section below 200 keV is insensitive to the NLD variation in the neutron channel, implying a domination of transitions to the g.s., it changes opposite to the NLD variation at higher energy. This can again be explained by the fact that the full neutron width is only included in the denominator of $\mathcal{F}$ when dealing with a laboratory cross section.

\subsection{Discussion of application}
\label{sec:examples}

The tables presented in the preceding sections contain the necessary information to assess uncertainties in astrophysical reaction rates.
The data presented are of importance also for the conception of theoretical and experimental studies to reduce said uncertainties.
Table \ref{tab:input} summarizes the conclusions from \S~\ref{sec:results} on what input quantities impact the widths.

As explained in the introduction, the error in the choice of models for both the reaction rate calculation and the input to the calculation cannot be quantified. Therefore it also depends on the type of input whether a systematic variation can be performed and what inherent uncertainty is assigned to the input. A further discussion goes beyond the scope of this work. Nevertheless, only a separation of the treatment of input uncertainties and rate sensitivities allows further insight in the dependences and their relevance. Although uncertainties in the input have to be determined separately, the sensitivities provided here are indispensable in determining the final impact of such uncertainties. Knowing that a certain input quantity $q$ carries a given uncertainty factor $u_q$, with the actual value $v_q$ of the quantity being in the range $v_q/u_q \leq v_q \leq u_qv_q$, Table \ref{tab:input} shows which width $c$ in the calculation is affected by it. This helps to estimate the uncertainty factor $u_c$ of $c$ which is often close to $u_q$ (except for masses). The sensitivities $^rs_c$ to a variation of that width can be found in Table \ref{tab:sensi}. The resulting uncertainty factor $\mathcal{U}$ in the astrophysical rate then is $\mathcal{U}=\; ^rs_cu_c$. In the case of a known systematic uncertainty factor $u_c^\mathrm{D}$ in the NLD in channel $c$, Table \ref{tab:nldsensi} can directly be used to obtain $^r_\mathrm{D}s_cu_c^\mathrm{D}$.

Practical applications will usually first determine the range of temperatures relevant to nuclear burning and a number of important reactions in a specific astrophysical process. Table \ref{tab:sensi} is then used to understand on which widths the astrophysical rate actually depends. The uncertainty in the relevant input and how it propagates to the width has to be estimated subsequently. Finally the resulting uncertainty in the rate is again computed with the sensitivity found in Table \ref{tab:sensi}.

Experimentalists interested in improving a specific rate, i.e., reducing its uncertainty, should also first note the plasma temperature range in which this reaction contributes to a given astrophysical process. The g.s.\ contributions $X$ in this temperature range, taken from Table \ref{tab:sensi}, allow to judge whether a measurement of the laboratory cross section actually can constrain the astrophysical reaction rate. This is only the case for $X$ being close to unity. In fact, the value of $X$ directly shows the maximally possible reduction of a rate uncertainty factor $\mathcal{U}\geq 1$. Assuming that the theoretical uncertainty factors of ground and excited state contributions are similar, $\mathcal{U}=\mathcal{U}_\mathrm{th}^\mathrm{gs}=\mathcal{U}_\mathrm{th}^\mathrm{exc}$, equation (18) of \citet{xfactor} can easily be rewritten to show the uncertainty factor $\mathcal{U}'$ remaining in the rate after a measurement,
\begin{equation}
\mathcal{U}'=\mathcal{U}_\mathrm{exp}^\mathrm{gs}X+\left(1-X\right)\mathcal{U}=\mathcal{U}_\mathrm{exp}^\mathrm{gs}+\left(1-X\right)\left(\mathcal{U}-\mathcal{U}_\mathrm{exp}^\mathrm{gs}\right) \quad,
\end{equation}
where $\mathcal{U}_\mathrm{exp}^\mathrm{gs}$ is the experimental uncertainty factor. A perfect measurement without any error bar would have $\mathcal{U}_\mathrm{exp}^\mathrm{gs}=1$, whereas a 30\% uncertainty, for example, would translate into $\mathcal{U}_\mathrm{exp}^\mathrm{gs}=1.3$.
The uncertainty in $X$ itself has been discussed in \citet{xfactor}. It scales inversely with the value of $X$ and has the property to keep large values of $X$ large and small ones small, thereby leaving unchanged any conclusions regarding the impact of a measurement.

Figures \ref{fig:Xng1p5}--\ref{fig:Xga1p5} show $X$ for a number of reactions. Values for other reactions and temperatures can be extracted from Tables \ref{tab:sensi}, \ref{tab:xnegq}, and \ref{tab:xphoto}. As can be seen, the values depend on temperature and NLD for a given reaction, not on separation from the line of stability. \citet{xfactor} have pointed out that even neutron captures on intermediate and heavy stable nuclei exhibit small $X$ at s-process temperatures and in consequence many s-process reactions are less well constrained by experiments than often assumed.

If $X\approx 1$, a measurement of the cross section would be the preferred way to proceed but it has to be performed in the energy range required for the calculation of the astrophysical rate. These energies are given in \citet{energywindows}. If it is not possible to measure at these energies, experiments can only indirectly -- with the involvement of theory -- improve a rate. If a measurement at higher energies is possible, Table \ref{tab:xstable} has to be consulted to check up to which energy the cross section is sensitive to the same widths as the rate. If the accessible cross sections exhibit different dependences, a cross section measurement is futile and alternative approaches have to be considered to improve the prediction of a rate. The alternative is to provide data to improve the calculation of the actually relevant widths. Table \ref{tab:input} shows which input is mainly relevant for which type of widths. Measurements can be performed to either directly improve the input or to provide data to test and improve the models used to calculate the input. Experimental determination of masses and nuclear spectroscopy for low-lying states are examples for the direct improvement, measurements of scattering and reaction cross section for a better determination of optical potentials are examples for providing local and global data for improving input to the calculation of the rate.

By examining Tables \ref{tab:sensi}--\ref{tab:nldxs}, theorists can directly see which nuclear properties affect an astrophysical rate most and may be in need of improvement. They can also immediately see the impact a different model prediction has on reaction rates and cross sections. One issue, already pointed out in \S~\ref{sec:intro}, should be remembered in this context. A similar procedure as suggested here for judging the sensitivities and resulting uncertainties in the rates has also to be applied to each input stemming from a theoretical model. The sensitivities of that model to its input have to be determined separately to be able to arrive at a systematic evaluation of the connected uncertainties. Comparing the results of a limited selection of models is not a systematic variation and might lead to incorrect conclusions when models fortuitously obtain similar values for astrophysically relevant quantities.

Sensitivity plots have already been used successfully in the interpretation of cross section measurements and their impact on astrophysical reaction rates \citep[see, e.g.,][]{gyurky06,gyurky07,kiss07,gyurky10,raureview,sauer11,hal12}. Such plots can be made using the data in Tables \ref{tab:sensi}--\ref{tab:nldxs}, they are also available online at \url{http://nucastro.org/nonsmoker.html}. Sample codes to extract plotting data for single reactions from the tables included here can be found at \url{http://download.nucastro.org/codes/}.

\section{Summary}
\label{sec:summary}

Sensitivities of astrophysical reaction rates to systematic variations of reaction widths and the NLD have been presented and general dependences have been discussed. Additionally, the sensitivities of reaction cross sections to the same quantities have been investigated. Considered were the compound nucleus reactions (n,$\gamma$), (n,p), (n,$\alpha$), (p,$\gamma$), (p,n), (p,$\alpha$), ($\alpha$,$\gamma$), ($\alpha$,n), ($\alpha$,p), ($\gamma$,n), ($\gamma$,p), and ($\gamma$,$\alpha$) for target nuclei between the driplines and $10\leq Z_A\leq 83$. General sensitivity trends have been discussed in detail and recommendations for application of the sensitivity information to assess and constrain uncertainties in astrophysical reaction rates were given.

The sensitivities and the shown trends are robust results which do not largely depend on the specific reaction code or the input used because they are calculated from ratios of quantities (rates or cross sections) and the basic dependence of the result on a given property is determined by fundamental nuclear properties and should be independent of the model code used. The only restriction is that a compound reaction mechanism is dominating the rate, either proceeding through narrow, discrete resonances or a large number of unresolved resonances. The (simpler) determination of sensitivities in the direct reaction mechanism has been outlined in \S~\ref{sec:reacmodels}.


%% Included in this acknowledgments section are examples of the
%% AASTeX hypertext markup commands. Use \url without the optional [HREF]
%% argument when you want to print the url directly in the text. Otherwise,
%% use either \url or \anchor, with the HREF as the first argument and the
%% text to be printed in the second.

\acknowledgments

This research was supported in part by the European Commission within the FP7 ENSAR/THEXO project and by the EuroGENESIS Collaborative Research Programme.

\clearpage

  \clearpage

  \clearpage

  \clearpage

  \clearpage

  \clearpage

  \clearpage

  \clearpage

  \clearpage

  \clearpage

  \clearpage

  \clearpage

  \clearpage

  \clearpage

  \clearpage

  \clearpage

  \clearpage

  \clearpage

  \clearpage

  \clearpage

  \clearpage

  \clearpage

  \clearpage

  \clearpage

  \clearpage

  \clearpage

%% If you are not including electonic art with your submission, you may
%% mark up your captions using the \figcaption command. See the
%% No more than seven \figcaption commands are allowed per page,
%% so if you have more than seven captions, insert a \clearpage

%% Tables should be submitted one per page, so put a \clearpage before

%% Two options are available to the author for producing tables:  the
%% deluxetable environment provided by the AASTeX package or the LaTeX
%% table environment.  Use of deluxetable is preferred.
%%

%% Three table samples follow, two marked up in the deluxetable environment,
%% one marked up as a LaTeX table.

%% In this first example, note that the \tabletypesize{}
%% command has been used to reduce the font size of the table.
%% We also use the \rotate command to rotate the table to
%% landscape orientation since it is very wide even at the
%% Note also that the \label command needs to be placed

%% This table also includes a table comment indicating that the full
%% version will be available in machine-readable format in the electronic

\clearpage

\begin{deluxetable}{rrrrrrrrrrcrrrrr}
\tabletypesize{\scriptsize}
\rotate
\tablecaption{Sensitivities of the astrophysical reaction rates to variations of different widths at 24 plasma temperatures; ground state contributions to the stellar rate are also given (see text for details)\label{tab:sensi}}
\tablewidth{0pt}
\tablehead{
\colhead{Element} & \colhead{$Z_A$} & \colhead{$M_A$} & \colhead{Proj.} & \colhead{Ejec.} & \colhead{$^\mathrm{r} s_\mathrm{g}$} & \colhead{$^\mathrm{r} s_\mathrm{n}$} & \colhead{$^\mathrm{r} s_\mathrm{p}$} & \colhead{$^\mathrm{r} s_\mathrm{a}$} & \colhead{$X$} & \colhead{\dots} & \colhead{$^\mathrm{r} s_\mathrm{g}$} & \colhead{$^\mathrm{r} s_\mathrm{n}$} & \colhead{$^\mathrm{r} s_\mathrm{p}$} & \colhead{$^\mathrm{r} s_\mathrm{a}$} & \colhead{$X$} \\
&&&&& \multicolumn{5}{c}{$T=0.1$ GK} &  & \multicolumn{5}{c}{$T=10.0$ GK}
}
\startdata
\dots &&&&&&&&&&&&&&& \\
mo & 42 & 92 & n & g & 0.56 & 0.35 & 0.00 & 0.00 & 1.000 & \dots & 0.87 & 0.07 & $-0.02$ & 0.00 & $1.02\times 10^{-1}$ \\
mo & 42 & 92 & n & p & $-0.37$ & 0.47 & 1.00 & 0.00 & 1.000 & \dots & $-0.01$ & 0.08 & $0.79$ & $-0.01$ & $6.57\times 10^{-4}$ \\
mo & 42 & 92 & n & a & $-0.25$ & 0.31 & 0.00 & 1.00 & 1.000 & \dots & 0.00 & 0.11 & $-0.15$ & 0.96 & $6.09\times 10^{-5}$ \\
mo & 42 & 92 & p & g & 0.00 & 0.00 & 1.00 & 0.00 & 1.000 & \dots & 1.00 & $-0.06$ & $0.09$ & 0.00 & $1.56\times 10^{-1}$ \\
mo & 42 & 92 & a & g & 0.00 & 0.00 & 0.00 & 1.00 & 1.000 & \dots & 0.65 & $-0.49$ & $-0.10$ & 0.96 & $3.86\times 10^{-1}$ \\
\dots &&&&&&&&&&&&&&&
\enddata
%% Text for table notes should follow after the \enddata but before
%% the \end{deluxetable}. Make sure there is at least one \tablenotemark
%% in the table for each \tablenotetext.
\tablecomments{Table \ref{tab:sensi} is published in its entirety in the
electronic edition of the {\it Astrophysical Journal Supplement}.  A portion is
shown here for guidance regarding its form and content.}
%\tablenotetext{a}{Sample footnote for table~\ref{tab:sensi} that was generated with the deluxetable environment}
%\tablenotetext{b}{Another sample footnote for table~\ref{tab:sensi}}
\end{deluxetable}

\clearpage

\begin{deluxetable}{rrrccllcrr}
\tabletypesize{\scriptsize}
%\rotate
\tablecaption{Ground state contributions $X$ to the stellar rate at 24 plasma temperatures for reactions with negative $Q$ value (except captures; see text for details)\label{tab:xnegq}}
\tablewidth{0pt}
\tablehead{
\colhead{Element} & \colhead{$Z_A$} & \colhead{$M_A$} & \colhead{Proj.} & \colhead{Ejec.} & \colhead{$X$}  & \colhead{$X$}  & \dots  & \colhead{$X$}  & \colhead{$X$} \\
&&&&& \colhead{$T=0.1$ GK} & \colhead{$T=0.15$} GK  & & \colhead{$T=9.0$ GK} & \colhead{$T=10.0$ GK}
}
\startdata
\dots &&&&&&&&& \\
mo & 42 & 92 & p & n & $3.73\times 10^{-6}$ & $3.73\times 10^{-6}$ & \dots & $2.53\times 10^{-3}$ & $1.39\times 10^{-3}$ \\
mo & 42 & 92 & p & a & $1.00\times 10^{0}$ & $1.00\times 10^{0}$ & \dots & $2.98\times 10^{-4}$ & $1.09\times 10^{-4}$ \\
mo & 42 & 92 & a & n & $1.00\times 10^{0}$ & $1.00\times 10^{0}$ & \dots & $6.27\times 10^{-2}$ & $2.63\times 10^{-2}$ \\
mo & 42 & 92 & a & p & $1.00\times 10^{0}$ & $9.98\times 10^{-1}$ & \dots & $3.25\times 10^{-2}$ & $1.17\times 10^{-2}$ \\
mo & 42 & 93 & p & n & $8.72\times 10^{-1}$ & $8.84\times 10^{-1}$ & \dots & $3.36\times 10^{-2}$ & $1.53\times 10^{-2}$ \\
\dots &&&&&&&&&
\enddata
%% Text for table notes should follow after the \enddata but before
%% the \end{deluxetable}. Make sure there is at least one \tablenotemark
%% in the table for each \tablenotetext.
\tablecomments{Table \ref{tab:xnegq} is published in its entirety in the
electronic edition of the {\it Astrophysical Journal Supplement}.  A portion is
shown here for guidance regarding its form and content.}
%\tablenotetext{a}{Sample footnote for table~\ref{tab:sensi} that was generated with the deluxetable environment}
%\tablenotetext{b}{Another sample footnote for table~\ref{tab:sensi}}
\end{deluxetable}

%\clearpage

\begin{deluxetable}{rrrcllcrr}
\tabletypesize{\scriptsize}
%\rotate
\tablecaption{Ground state contributions $X$ to the stellar rate at 24 plasma temperatures for photodisintegration reactions (see text for details)\label{tab:xphoto}}
\tablewidth{0pt}
\tablehead{
\colhead{Element} & \colhead{$Z_A$} & \colhead{$M_A$} & \colhead{Ejec.} & \colhead{$X$}  & \colhead{$X$}  & \dots  & \colhead{$X$}  & \colhead{$X$} \\
&&&& \colhead{$T=0.1$ GK} & \colhead{$T=0.15$} GK  & & \colhead{$T=9.0$ GK} & \colhead{$T=10.0$ GK}
}
\startdata
\dots &&&&&&&& \\
mo & 42 & 92 & n & $5.93\times 10^{-7}$ & $5.93\times 10^{-7}$ & \dots & $4.14\times 10^{-4}$ & $3.32\times 10^{-4}$ \\
mo & 42 & 92 & p & $2.03\times 10^{-4}$ & $2.03\times 10^{-4}$ & \dots & $2.09\times 10^{-3}$ & $1.56\times 10^{-3}$ \\
mo & 42 & 92 & a & $1.00\times 10^{0}$ & $1.43\times 10^{-1}$ & \dots & $1.48\times 10^{-3}$ & $6.36\times 10^{-4}$ \\
mo & 42 & 93 & n & $3.92\times 10^{-2}$ & $3.92\times 10^{-2}$ & \dots & $3.13\times 10^{-3}$ & $2.14\times 10^{-3}$ \\
mo & 42 & 93 & p & $2.15\times 10^{-5}$ & $2.15\times 10^{-5}$ & \dots & $3.50\times 10^{-4}$ & $2.69\times 10^{-4}$ \\
\dots &&&&&&&&
\enddata
%% Text for table notes should follow after the \enddata but before
%% the \end{deluxetable}. Make sure there is at least one \tablenotemark
%% in the table for each \tablenotetext.
\tablecomments{Table \ref{tab:xphoto} is published in its entirety in the
electronic edition of the {\it Astrophysical Journal Supplement}.  A portion is
shown here for guidance regarding its form and content.}
%\tablenotetext{a}{Sample footnote for table~\ref{tab:sensi} that was generated with the deluxetable environment}
%\tablenotetext{b}{Another sample footnote for table~\ref{tab:sensi}}
\end{deluxetable}

\clearpage

\begin{deluxetable}{rrrrrrrrrrrcrrrrr}
\tabletypesize{\scriptsize}
\rotate
\tablecaption{Sensitivities of laboratory cross sections to variations of different widths (see text for details)\label{tab:xstable}}
\tablewidth{0pt}
\tablehead{
\colhead{Element} & \colhead{$Z_A$} & \colhead{$M_A$} & \colhead{Proj.} & \colhead{Ejec.} & \colhead{$n_\mathrm{En}$} & \colhead{$E$ (MeV)} & \colhead{$^\mathrm{cs} s_\mathrm{g}$} & \colhead{$^\mathrm{cs} s_\mathrm{n}$} & \colhead{$^\mathrm{cs} s_\mathrm{p}$} & \colhead{$^\mathrm{cs} s_\mathrm{a}$} & \colhead{\dots} & \colhead{$E$ (MeV)} & \colhead{$^\mathrm{cs} s_\mathrm{g}$} & \colhead{$^\mathrm{cs} s_\mathrm{n}$} & \colhead{$^\mathrm{cs} s_\mathrm{p}$} & \colhead{$^\mathrm{cs} s_\mathrm{a}$}
}
\startdata
\dots &&&&&&&&&&&&&&&& \\
mo & 42 & 92 & n & g & 50 & 0.001 & 0.44 & 0.49 & 0.00 & 0.00 & \dots & 14.131 & 0.99 & 0.15 & $-0.22$ & $-0.03$ \\
mo & 42 & 92 & n & p & 50 & 0.001 & $-0.63$ & 0.74 & 1.00 & 0.00 & \dots & 14.131 & 0.00 & 0.15 & $0.64$ & $-0.03$ \\
mo & 42 & 92 & n & a & 50 & 0.001 & $-0.30$ & 0.37 & 0.00 & 1.00 & \dots & 14.131 & 0.00 & 0.15 & $-0.22$ & 0.95 \\
mo & 42 & 92 & p & g & 50 & 0.270 & 0.00 & 0.00 & 1.00 & 0.00 & \dots & 7.141 & 1.06 & 0.00 & 0.00 & 0.00 \\
mo & 42 & 92 & p & n & 50 & 8.626 & 0.00 & 0.00 & 0.00 & 0.00 & \dots & 14.141 & $-0.01$ & 0.21 & 0.52 & 0.00 \\
\dots &&&&&&&&&&&&&&&&
\enddata
%% Text for table notes should follow after the \enddata but before
%% the \end{deluxetable}. Make sure there is at least one \tablenotemark
%% in the table for each \tablenotetext.
\tablecomments{Table \ref{tab:xstable} is published in its entirety in the
electronic edition of the {\it Astrophysical Journal Supplement}.  A portion is
shown here for guidance regarding its form and content.}
%\tablenotetext{a}{Sample footnote for table~\ref{tab:sensi} that was generated with the deluxetable environment}
%\tablenotetext{b}{Another sample footnote for table~\ref{tab:sensi}}
\end{deluxetable}

%\clearpage

\begin{deluxetable}{rrrrrrrrrrrrrrrrrc}
\tabletypesize{\scriptsize}
\rotate
\tablecaption{Sensitivities of the astrophysical reaction rates to variations of nuclear level densities (with and without the inclusion of discrete excited states) in the $\gamma$-, neutron-, proton-, and $\alpha$-channel at 24 plasma temperatures (see text for details)\label{tab:nldsensi}}
\tablewidth{0pt}
\tablehead{
\colhead{Element} & \colhead{$Z_A$} & \colhead{$M_A$} & \colhead{Proj.} & \colhead{Ejec.} & \colhead{$E_\mathrm{last}^\mathrm{g}$} & \colhead{$E_\mathrm{last}^\mathrm{n}$} & \colhead{$E_\mathrm{last}^\mathrm{p}$} & \colhead{$E_\mathrm{last}^\mathrm{a}$} &
\colhead{$^\mathrm{r} _\mathrm{D} s_\mathrm{g}$} & \colhead{$^\mathrm{r} _\mathrm{D} s_\mathrm{n}$} & \colhead{$^\mathrm{r} _\mathrm{D} s_\mathrm{p}$} & \colhead{$^\mathrm{r} _\mathrm{D} s_\mathrm{a}$} & \colhead{$^\mathrm{r} _\mathrm{D} s_\mathrm{g}^0$} & \colhead{$^\mathrm{r} _\mathrm{D} s_\mathrm{n}^0$} & \colhead{$^\mathrm{r} _\mathrm{D} s_\mathrm{p}^0$} & \colhead{$^\mathrm{r} _\mathrm{D} s_\mathrm{a}^0$} & \dots \\
&&&&&&&&& \multicolumn{8}{c}{$T=0.1$ GK} &
}
\startdata
\dots &&&&&&&&&&&&&&&&& \\
mo & 42 & 92 & n & g & 2.77 & 5.12 & 0.501 & 4.10 & 0.46 & 0.00 & 0.00 & 0.00 & 0.55 & 0.00 & 0.00 & 0.00 & \dots \\
mo & 42 & 92 & n & p & 2.77 & 5.12 & 0.501 & 4.10 & $-0.31$ & 0.00 & 0.00 & 0.00 & $-0.11$ & 0.00 & 1.00 & 0.00 & \dots \\
mo & 42 & 92 & n & a & 2.77 & 5.12 & 0.501 & 4.10 & $-0.20$ & 0.00 & 0.00 & 0.00 & $-0.24$ & 0.00 & 0.00 & 0.01 & \dots \\
mo & 42 & 92 & p & g & 1.79 & 0.966 & 5.12 & 0.659 & 0.00 & 0.00 & 0.00 & 0.00 & 0.00 & 0.00 & 0.00 & 0.00 & \dots \\
mo & 42 & 92 & a & g & 3.95 & 1.35 & 2.23 & 5.12 & 0.00 & 0.00 & 0.00 & 0.00 & 0.00 & 0.00 & 0.00 & 0.00 & \dots \\
\dots &&&&&&&&&&&&&&&&&
\enddata
%% Text for table notes should follow after the \enddata but before
%% the \end{deluxetable}. Make sure there is at least one \tablenotemark
%% in the table for each \tablenotetext.
\tablecomments{Table \ref{tab:nldsensi} is published in its entirety in the
electronic edition of the {\it Astrophysical Journal Supplement}.  A portion is
shown here for guidance regarding its form and content.}
%\tablenotetext{a}{Sample footnote for table~\ref{tab:sensi} that was generated with the deluxetable environment}
%\tablenotetext{b}{Another sample footnote for table~\ref{tab:sensi}}
\end{deluxetable}

\begin{deluxetable}{rrrrrrrrrrrrrrrc}
\tabletypesize{\scriptsize}
\rotate
\tablecaption{Sensitivities of laboratory cross sections to variations of nuclear level densities in the $\gamma$-, neutron-, proton-, and $\alpha$-channel, with and without inclusion of discrete excited states (see text for details)\label{tab:nldxs}}
\tablewidth{0pt}
\tablehead{
\colhead{Element} & \colhead{$Z_A$} & \colhead{$M_A$} & \colhead{Proj.} & \colhead{Ejec.} & \colhead{$n_\mathrm{En}$} & \colhead{$E$ (MeV)} & \colhead{$^\mathrm{cs} _\mathrm{D} s_\mathrm{g}$} & \colhead{$^\mathrm{cs} _\mathrm{D} s_\mathrm{n}$} & \colhead{$^\mathrm{cs} _\mathrm{D} s_\mathrm{p}$} & \colhead{$^\mathrm{cs} _\mathrm{D} s_\mathrm{a}$} & \colhead{$^\mathrm{cs} _\mathrm{D} s^0_\mathrm{g}$} & \colhead{$^\mathrm{cs} _\mathrm{D} s^0_\mathrm{n}$} & \colhead{$^\mathrm{cs} _\mathrm{D} s^0_\mathrm{p}$} & \colhead{$^\mathrm{cs} _\mathrm{D} s^0_\mathrm{a}$} & \colhead{\dots}
}
\startdata
\dots &&&&&&&&&&&&&&& \\
mo & 42 & 92 & n & g & 50 & 0.001 & 0.43 & 0.00 & 0.00 & 0.00 & 0.51 & 0.00 & 0.00 & 0.00 & \dots \\
mo & 42 & 92 & n & p & 50 & 0.001 & $-0.45$ & 0.00 & 0.00 & 0.00 & $-0.21$ & 0.00 & 1.00 & 0.00 & \dots \\
mo & 42 & 92 & n & a & 50 & 0.001 & $-0.22$ & 0.00 & 0.00 & 0.00 & $-0.27$ & 0.00 & 0.00 & 0.01 & \dots \\
mo & 42 & 92 & p & g & 50 & 1.378 & 0.00 & 0.00 & 0.00 & 0.00 & 0.00 & 0.00 & 0.00 & 0.00 & \dots \\
mo & 42 & 92 & p & n & 50 & 8.626 & 0.00 & 0.00 & 0.00 & 0.00 & 0.00 & 0.00 & 0.00 & 0.00 & \dots \\
\dots &&&&&&&&&&&&&&&
\enddata
%% Text for table notes should follow after the \enddata but before
%% the \end{deluxetable}. Make sure there is at least one \tablenotemark
%% in the table for each \tablenotetext.
\tablecomments{Table \ref{tab:nldxs} is published in its entirety in the
electronic edition of the {\it Astrophysical Journal Supplement}.  A portion is
shown here for guidance regarding its form and content.}
%\tablenotetext{a}{Sample footnote for table~\ref{tab:sensi} that was generated with the deluxetable environment}
%\tablenotetext{b}{Another sample footnote for table~\ref{tab:sensi}}
\end{deluxetable}

\clearpage

\begin{deluxetable}{cc}
%\tabletypesize{\scriptsize}
%\setlength{\tabcolsep}{0.02in}
\tablecaption{Input to the compound reaction calculation and its impact (in
no particular order).\label{tab:input}}
\tablewidth{0pt}
\tablehead{
\colhead{Quantity} & \colhead{Impact}
}
\startdata
Nuclear masses & all widths\tablenotemark{a} \\
Properties of g.s.\ and excited states & particle widths\tablenotemark{b}\\
Nuclear level density & $\gamma$-width\tablenotemark{b}\\
Optical potentials & particle widths\tablenotemark{c}\\
$\gamma$ strength function & $\gamma$-widths\tablenotemark{d}\\
\tableline
Deformations & optical potentials, NLD, $\gamma$ strength function\\
(Nuclear matter density distribution) & (optical potentials)\tablenotemark{e}
\enddata
\tablenotetext{a}{Determining the relative energy in all transitions through separation energies}
\tablenotetext{b}{In principle, all widths are affected by inclusion of discrete levels up to an energy $E_\mathrm{last}$, with specific excitation energy, spin, parity, and the use of a NLD above, see Equation (\ref{eq:transcoeff}). Relevant $\gamma$-energies, however, are $2-4$ MeV and mainly lead to states above $E_\mathrm{last}$ whereas particle transitions prefer low-lying states.}
\tablenotetext{c}{Separate optical potentials per particle type}
\tablenotetext{d}{E1 is dominating, depending on NLD sometimes also M1 relevant}
\tablenotetext{e}{Some optical potential descriptions require the knowledge of nuclear densities, e.g., folding potentials or microscopic potentials based on the local density approximation \citep[e.g.,][]{jlm,bauge}}
\end{deluxetable}

\end{document}